\documentclass[aps]{revtex4}
\usepackage{eurosym}
\usepackage{amsfonts}
\usepackage{amsmath}
\usepackage{amssymb,epsf}
\usepackage{color}
\usepackage{graphicx}
\usepackage{epstopdf}
\usepackage{float}
\usepackage{caption}
\usepackage{subfig}
\usepackage{comment}
\usepackage{ulem}

\begin{document}

\title{Accelerating AdS black holes in gravity's rainbow}
\author{ B. Eslam Panah$^{1,2,3}$ \footnote{%
email address: eslampanah@umz.ac.ir}, S. Zare$^{5}$ \footnote{
email address: soroushzrg@gmail.com}, and H. Hassanabadi $^{4,5}$ \footnote{%
email address: hha1349@gmail.com } }
\affiliation{$^{1}$ Department of Theoretical Physics, Faculty of Basic Sciences, University of Mazandaran, P. O. Box 47416-95447, Babolsar, Iran\\
$^{2}$ ICRANet-Mazandaran, University of Mazandaran, P. O. Box 47415-416,
Babolsar, Iran\\
$^{3}$ ICRANet, Piazza della Repubblica 10, I-65122 Pescara, Italy\\
$^{4}$ Faculty of Physics, Shahrood University of Technology, Shahrood, Iran\\
$^{5}$ Department of Physics, University of Hradec Kr\'{a}lov\'{e}, Rokitansk%
\'{e}ho 62, 500 03 Hradec Kr\'{a}lov\'{e}, Czechia }

\begin{abstract}
Motivated by the effect of the energy of moving particles in $C-$metric, we
first obtain exact accelerating black hole solutions in gravity's rainbow.
Then, we study the effects of gravity's rainbow and $C-$metric parameters on
the Ricci and Kretschmann scalars, and also the asymptotical behavior of
this solution. Next, we indicate how different parameters of the obtained
accelerating black holes in gravity's rainbow affect thermodynamics
quantities (such as the Hawking temperature, and entropy) and the local
stability (by evaluating the heat capacity). In the following, we extract
the geodesic equations to determine the effects of various parameters on
photon trajectory in the vicinity of this black hole, as well as obtain the
radius of the photon sphere and the corresponding critical impact parameter
to gain insight into AdS black hole physics by adding the gravity's rainbow
to $C-$metric.
\end{abstract}

\maketitle

\section{\textbf{Introduction}}

Black holes are some of the most fascinating and mind-bending objects in the
cosmos. They can help our knowledge from the points of theoretical and
experimental theories of physics. Among them, one of the most interesting
black holes is related to the accelerating black hole, which is described by
the $C-$metric \cite{Acc1,Acc2,Acc3,Acc4,Acc5}. The accelerating black hole
includes a conical singularity which can be imagined as a cosmic string with
a tension providing the force driving the acceleration. This black hole
attracted much attention due to the existence of a string-like singularity
along one polar axis attached to it \cite%
{BHacc1,BHacc2,BHacc3,BHacc4,BHacc5,BHacc6,BHacc7,BHacc8,BHqcc8b,BHacc9}. In
addition, there has been a significant amount of research conducted on
various aspects related to the accelerating black holes. These include the
examination of their global causal structure \cite{causal}, black hole
shadows \cite{Zhang2021}, quantum thermal properties \cite{Yu1995},
holographic heat engines and complexity \cite%
{Zhang2018,Zhang2019,Jiang2021,Zhang2023}, and more. One specific area of
focus has been the study of the thermodynamics of accelerating black holes,
which the first time was explored in Ref. \cite{BHacc1} and further
addressed in subsequent references \cite%
{BHacc2,BHacc4,Gregory2019,Ball2021,Cassani2021}. These studies have
successfully extended the first law of thermodynamics, the Bekentein-Smarr
and Christodoulou-Ruffini-type mass formula, to encompass (un)charged
rotating accelerating black holes.

Studying the effects of modified theories of gravity on the accelerating
black hole's properties is an interesting subject. For example, three and
four-dimensional accelerating black holes in $F(R)$ gravity have been
evaluated \cite{F(R)acc1} and \cite{F(R)acc2}, respectively. In addition, a
three-dimensional accelerating black hole in gravity's rainbow is obtained
in Ref. \cite{3accBH}. However, there are no four-dimensional accelerating
black hole solutions in gravity's rainbow yet. So, we first focus on
extracting the accelerating black hole solutions in this theory of gravity.

Gravity's rainbow, initially proposed by Magueijo and Smolin and
investigated within the framework of double special relativity \cite%
{MagueijoCQG2004-1}, introduces a geometry dependent on the energy of moving
particles. In this formalism, varying particle energies lead to distinct
distortions in spacetime. When investigating the quantum gravity effects of
moving probes on the geometry, the notion of a singular spacetime background
is replaced by a family of line elements, referred to as the ``rainbow
functions" which are parameterized by the energy of these moving probes.
Gravity's rainbow theory admitted the invariant energy scale connected to
the Planck energy and the invariant velocity of light at low energies \cite%
{Amelino-CameliaPLB2001,Amelino-CameliaIJMPD2002,Amelino-CameliaIJMPA2005,MagueijoPRL2002,MagueijoPRD2003,GalanPRD2004,deMontigny}%
. The modified energy-momentum dispersion relation in this formalism can be
expressed as follows \cite%
{MagueijoPRL2002,MagueijoPRD2003,MagueijoCQG2004,DarllaUniv2023} 
\begin{equation}
E^{2}F^{2}\left( {\varepsilon }\right) -p^{2}c^{2}H^{2}\left( {\varepsilon }%
\right) =m^{2}c^{4}  \label{MEMDR}
\end{equation}%
where $F\left( {\varepsilon }\right) $ and $H\left( {\varepsilon }\right) $
stand for rainbow functions that are characterized by the ratio {$%
\varepsilon =$}$\frac{E}{E_{p}}$. Here, $E_{p}$ denotes the energy on the
Planck scale, while $E$ represents the energy of the system. To investigate
the impact of the gravity's rainbow model on the thermodynamics and geodesic
equations for this black hole -- which we will subsequently utilize -- the
following rainbow functions are taken into account \cite%
{AwadJCAP2013,AliPRD2014,AshourEPJC2016} 
\begin{equation}
F\left( {\varepsilon }\right) =1,{~~}\&~~\ H\left( {\varepsilon }\right) =%
\sqrt{1-\gamma {\varepsilon }^{2}},  \label{RF1}
\end{equation}
here, $\gamma $ denotes a dimensionless free parameter in the model.

In light of gravity's rainbow theory and its applications in the literature,
people have explored theories of cosmology, astrophysics, modified gravity,
as well as deformed structure of the spacetime around the black hole region
and wormhole geometries \cite%
{MajumderIJMPD2013,NeutronS1,NeutronS2,DarkS1,DarkS2,GravaS1,GravaS2,HendiEPJC2016,LeivaMPLA2009,Ling2007,Awad2013,HendiPoDU2017,BezerraEpL2017,GarattiniEPJC2015,GarattiniPLB2010,GarattiniPRD2011,GarattiniPRD2012,GarattiniNPB2014,GarattiniPRD2012-2,GarattiniPRD2014,AliIJGMMP2015}%
. Following the release of the EHT images \cite{KocherlakotaPRD2020},
scientists became interested in comparing the data from EHT with theoretical
models to determine the black hole's characteristics, including mass and
spin \cite%
{CapozzielloJCAP2023,Capozziello2023,PerlickPR2022,GrallaPRD2019,PengCPC2021,UniyalPoDU2023}%
. A clear picture of how the acceleration of the black hole impacts the
photon trajectory around the AdS black hole in energy-dependent $C-$metric
is still a mystery, and in this paper, we focus on one such case. Therefore,
we are interested in investigating the trajectories of light rays in the
vicinity of such an accelerating black hole, as well as determine the radius
of the photon sphere and the corresponding critical impact parameter, to
gain insight into AdS black hole physics by adding the gravity's rainbow to $%
C-$metric.

\section{\textbf{Exact Solutions}}

To obtain the accelerating black hole, we have to construct a kind of
energy-dependent $C-$metric. For this purpose we use the mentioned method in
Ref. \cite{Peng2008} as 
\begin{equation}
h\left( \varepsilon \right) =\eta ^{\mu \nu }e_{\mu }\left( \varepsilon
\right) \otimes e_{\nu }\left( \varepsilon \right) ,
\end{equation}%
where $e_{0}\left( \varepsilon \right) =\frac{1}{F\left( \varepsilon \right) 
}\widetilde{e_{0}}$, and $e_{i}\left( \varepsilon \right) =\frac{1}{H\left(
\varepsilon \right) }\widetilde{e_{i}}$. Notably, the tilde quantities refer
to the energy-independent frame fields. Using the above conditions, we can
create a suitable energy-dependent $C-$metric to obtain an accelerating
black hole in gravity's rainbow. Considering the introduced $C-$metric in
Refs. \cite{BHacc1} and by applying $e_{0}\left( \varepsilon \right) $ and $%
e_{i}\left( \varepsilon \right) $, we can get energy-dependent $C-$metric as 
\begin{equation}
ds^{2}=\frac{1}{\mathcal{K}^{2}\left( r,\theta \right) }\left[ -\frac{f(r)}{%
F^{2}\left( \varepsilon \right) }dt^{2}+\frac{dr^{2}}{f(r)H^{2}\left(
\varepsilon \right) }+\frac{r^{2}}{H^{2}\left( \varepsilon \right) }\left( 
\frac{d\theta ^{2}}{g\left( \theta \right) }+\frac{g\left( \theta \right)
\sin ^{2}\theta d\varphi ^{2}}{K^{2}}\right) \right] ,  \label{Metric}
\end{equation}%
where $\mathcal{K}\left( r,\theta \right) =1+Ar\cos \theta $, which is
called the conformal factor.

Now, we are in a position to find suitable metric functions $f(r)$ and $%
g\left( \theta \right) $ by using all components of equations of motion $%
G_{\mu \nu }+\Lambda \mathrm{g}_{\mu \nu }=0$ (where $\mathrm{g}_{\mu \nu }$
is metric tensor). Considering the metric (\ref{Metric}) and equations of
motion, one can find that 
\begin{eqnarray}
Eq_{tt} &=&Eq_{rr}=\sin \theta g^{\mathcal{\theta \theta }}\mathcal{K}%
^{2}\left( r,\theta \right) +3\left( Ar\cos ^{2}\theta +\frac{4Ar\sin
^{2}\theta }{3}+\cos \theta \right) g^{\mathcal{\theta }}\mathcal{K}\left(
r,\theta \right)  \notag \\
&&  \notag \\
&&+2r^{2}\sin \theta \left[ 3gA^{2}\cos ^{2}\theta +\frac{\left( f-g\right)
\left( 1-2Ar\cos \theta \right) }{r^{2}}+\frac{\mathcal{K}\left( r,\theta
\right) f^{\prime }}{r}+\frac{\Lambda }{H^{2}\left( \varepsilon \right) }%
\right] ,  \label{tt1} \\
&&  \notag \\
Eq_{\theta \theta } &=&Eq_{\varphi \varphi }=r^{2}\mathcal{K}^{2}\left(
r,\theta \right) f^{\prime \prime }+2\left( 1-A^{2}r^{2}\cos ^{2}\theta
\right) rf^{\prime }+2Ar\sin \theta \mathcal{K}\left( r,\theta \right) g^{%
\mathcal{\theta }}  \notag \\
&&  \notag \\
&&+6gA^{2}r^{2}\sin ^{2}\theta +\frac{2\Lambda r^{2}}{H^{2}\left(
\varepsilon \right) }+2Ar\cos \theta \left[ 2\left( g-f\right) +Ar\cos
\theta \left( f+2g\right) \right] ,  \label{phiphi1}
\end{eqnarray}%
where $f=f\left( r\right) $, $g=g\left( \theta \right) $, $f^{\prime }=\frac{%
df(r)}{dr}$, $f^{\prime \prime }=\frac{d^{2}f(r)}{dr^{2}}$, $g^{\mathcal{%
\theta }}=\frac{dg(\theta )}{d\theta }$, and $g^{\mathcal{\theta \theta }}=%
\frac{d^{2}g(\theta )}{d\theta ^{2}}$. It is notable that $Eq_{tt}$, $%
Eq_{rr} $, $Eq_{\theta \theta }$ and $Eq_{\varphi \varphi }$ are related to
components of $tt$, $rr$, $\theta \theta $\ and $\varphi \varphi $ of the
equations of motion.

After some calculations, we find the exact solutions of Eqs. (\ref{tt1}) and
(\ref{phiphi1}) for the functions $f(r)$ and $g\left( \theta \right) $ in
the following forms 
\begin{eqnarray}
f\left( r\right) &=&\left( 1-A^{2}r^{2}\right) \left( 1-\frac{2m}{r}\right) -%
\frac{\Lambda r^{2}}{3H^{2}\left( \varepsilon \right) },  \notag \\
&&  \notag \\
g\left( \theta \right) &=&1+2mA\cos \theta ,  \label{f(r)Uch}
\end{eqnarray}%
where $\Lambda $, and $m$, are the cosmological constant, and a constant
that is related to the total mass of the black hole, respectively. It is
worthwhile to mention that we consider $G=c=1$.

Notably, we can define $K$ as introduced in Refs. \cite{BHacc1,BHacc4},
which is related to the presence of cosmic string. In other words, by
looking at the angular part of the metric and the behavior of $g\left(\theta
\right) $ at both poles $\theta _{+}=0$ (north pole), and $\theta_{-}=\pi $
(south pole), we can find the presence of cosmic string. The regularity of
the metric at a pole requires $K_{\pm }=g\left( \theta _{\pm}\right) =1\pm
2mA$, where $K_{\pm }$ is chosen to regularize one pole and another pole is
left with either a conical deficit or a conical excess along the other pole.
Here we would make the black hole regular on the north pole, i.e., $\theta
=0 $, by fixing $K=K_{+}=1+2mA$.

Also, in the absence of the accelerating parameter ($A=0$), the solution (%
\ref{f(r)Uch}) reduces to the black hole solutions in gravity's rainbow in
the form 
\begin{equation}
f\left( r\right) =1-\frac{2m}{r}-\frac{\Lambda r^{2}}{3H^{2}\left(
\varepsilon \right) },
\end{equation}%
and by considering $H^{2}\left( \varepsilon \right) =1$, the above solution
turns to the well-known Schwartzshield black hole solutions in the present
of the cosmological constant as $f\left( r\right) =1-\frac{2m}{r}-\frac{%
\Lambda r^{2}}{3}$.

To find the curvature singularity(ies) of the spacetime, we calculate the
Ricci and Kretschmann scalars. Using the metric (\ref{Metric}) and after
some algebraic manipulation, one can find the Ricci scalar ($\mathcal{R}$)
and Kretschmann scalar ($\mathcal{R}_{\alpha \beta \gamma \delta }\mathcal{R}%
^{\alpha \beta \gamma \delta }$) in the following forms 
\begin{eqnarray}
\mathcal{R} &=&4\Lambda , \\
&&  \notag \\
\mathcal{R}_{\alpha \beta \gamma \delta }\mathcal{R}^{\alpha \beta \gamma
\delta } &=&\frac{48\mathcal{A}_{0}}{r^{6}}+\frac{288\mathcal{A}_{1}}{r^{5}}+%
\frac{720\mathcal{A}_{2}}{r^{4}}+\frac{960\mathcal{A}_{3}}{r^{3}}+\frac{720%
\mathcal{A}_{4}}{r^{2}}+\frac{288\mathcal{A}_{5}}{r}  \notag \\
&&+48\mathcal{A}_{6}+\frac{8\Lambda ^{2}}{3},
\end{eqnarray}%
where $\mathcal{A}_{n}=H^{4}\left( \varepsilon \right) m^{2}A^{n}\cos
^{n}\theta $. The above equation indicates that the Kretschmann scalar
diverges at $r=0$, i.e.,%
\begin{equation}
\underset{r\longrightarrow 0}{\lim }\mathcal{R}_{\alpha \beta \gamma \delta }%
\mathcal{R}^{\alpha \beta \gamma \delta }\longrightarrow \infty ,
\end{equation}%
so we encounter with a curvature singularity at $r=0$. Also, the
asymptotical behavior is dependent on the parameters of this theory. Indeed,
the rainbow function $H\left( \varepsilon \right) $ and the parameter of
accelerating black holes affect the asymptotical behavior of spacetime, i.e.,%
\begin{equation}
\underset{r\longrightarrow \infty }{\lim }\mathcal{R}_{\alpha \beta \gamma
\delta }\mathcal{R}^{\alpha \beta \gamma \delta }\longrightarrow 48\mathcal{A%
}_{6}+\frac{8\Lambda ^{2}}{3}.
\end{equation}

\begin{figure}[tbph]
\centering
\includegraphics[width=0.3\linewidth]{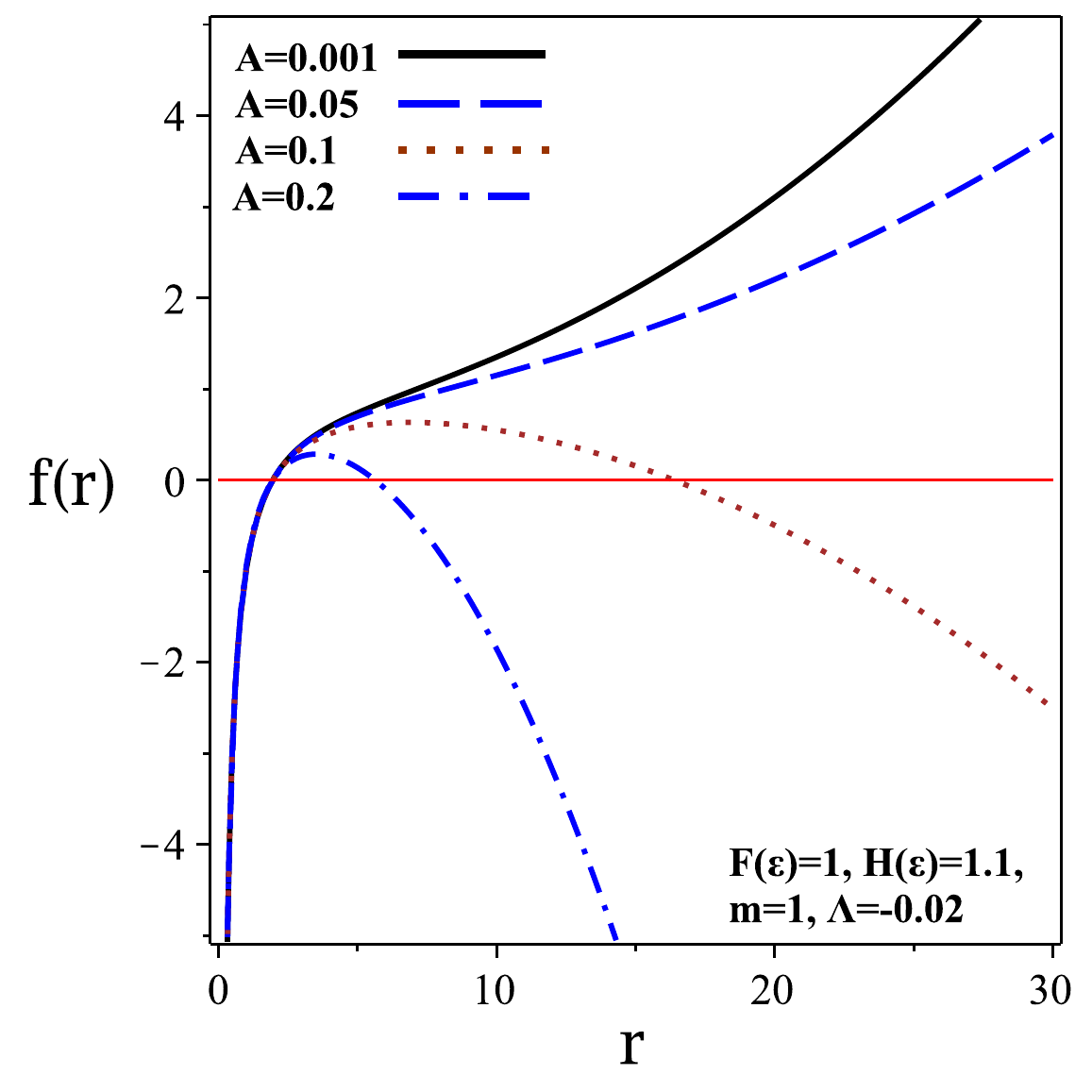} \includegraphics[width=0.3%
\linewidth]{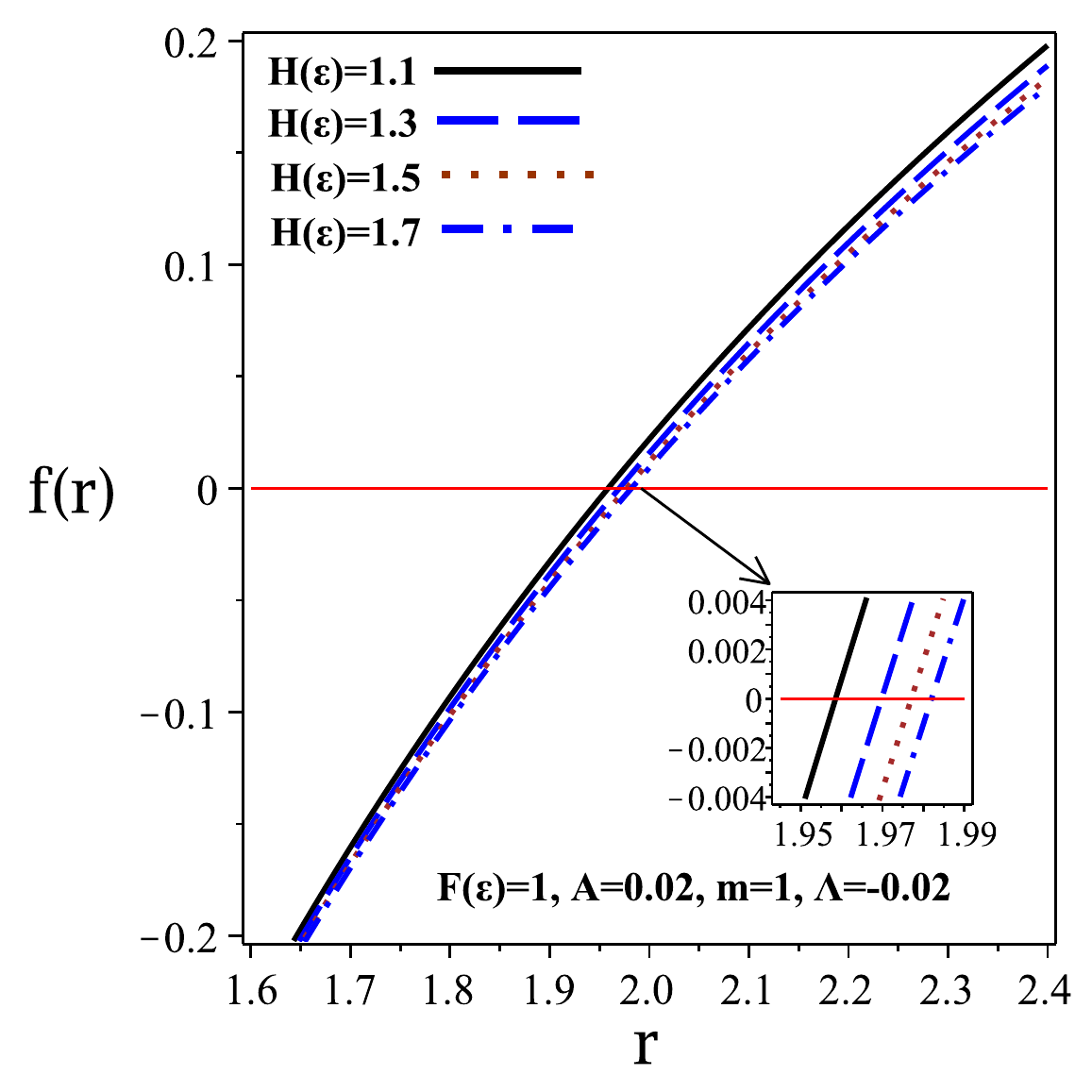} \newline
\includegraphics[width=0.3\linewidth]{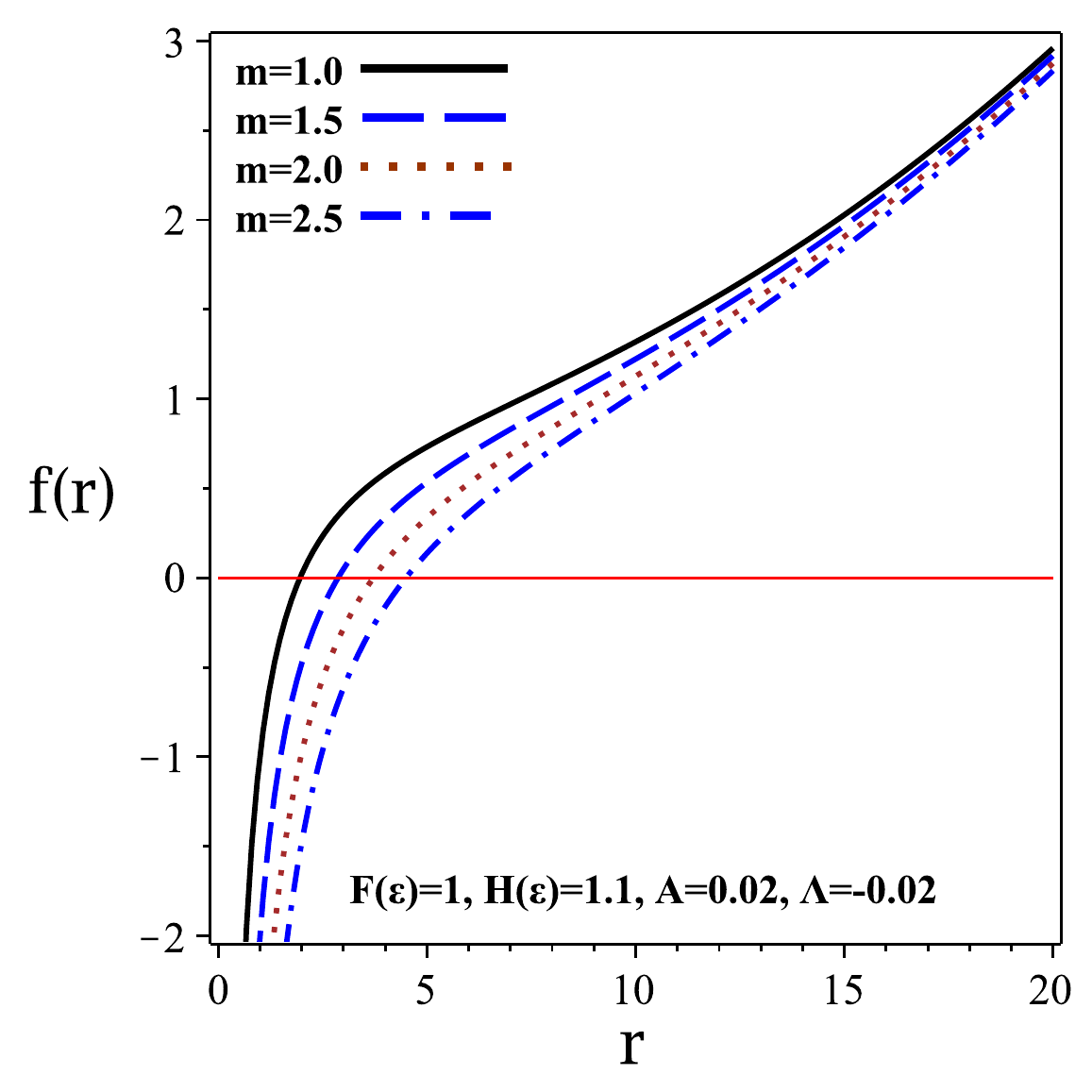} \includegraphics[width=0.3%
\linewidth]{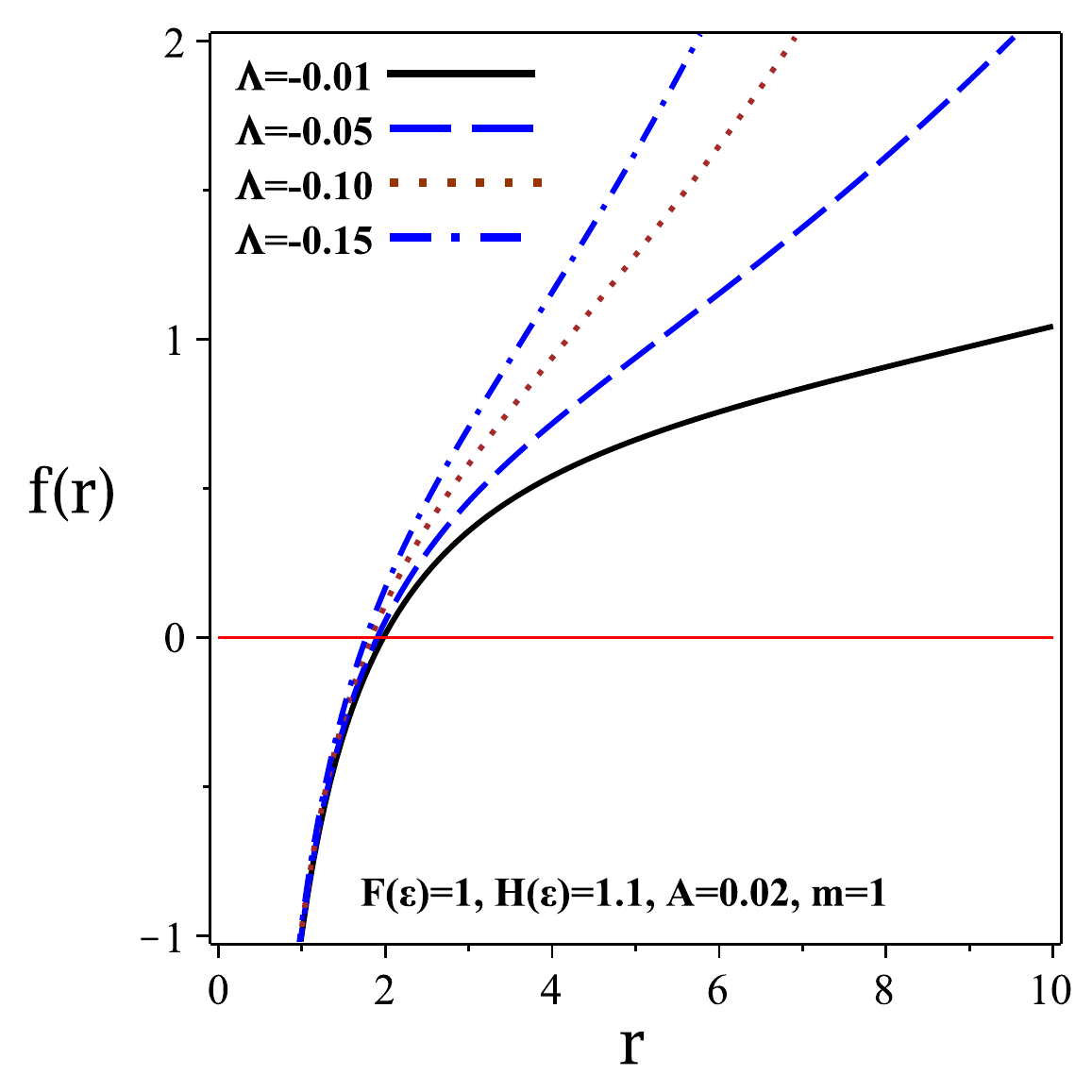} \newline
\caption{$f(r)$ versus $r$ for different values of parameters.}
\label{Fig1}
\end{figure}

We study the effects of various parameters on horizons. Our findings
indicate that; i) by increasing the accelerating parameter, these black
holes may encounter with two horizons (inner and outer horizons), see the up
left panel in Fig. \ref{Fig1}. ii) by increasing the rainbow function $%
H\left( \varepsilon \right) $, the event horizon increases (see the up right
panel in Fig. \ref{Fig1}). iii) massive black holes have large radii, as we
expected (see the down left panel in Fig. \ref{Fig1}). iv) by increasing $%
\left\vert \Lambda \right\vert $, we encounter with small black holes (see
the down right panel in Fig. \ref{Fig1}).

\section{\textbf{Thermodynamics}}

Considering the black hole as a thermodynamic system, we are going to obtain
some of the conserved and thermodynamic quantities of the accelerating black
holes in the content of gravity's rainbow such as the Hawking temperature,
entropy and then study the local stability by evaluating the heat capacity.

\subsection{\textbf{Hawking Temperature}}

By equating $g_{tt}=f(r)=0$, we get the geometrical mass ($m$) in the
following form 
\begin{equation}
m=\frac{(\Lambda +3A^{2}H^{2}(\varepsilon ))r_{+}^{3}-3H^{2}(\varepsilon
)r_{+}}{6(A^{2}r_{+}^{2}-1)H^{2}(\varepsilon )}.  \label{m}
\end{equation}

To get the Hawking temperature, we employ the definition of surface gravity 
\begin{equation}
\kappa =\sqrt{\frac{-1}{2}\left( \nabla _{\mu }\chi _{\nu }\right) \left(
\nabla ^{\mu }\chi ^{\nu }\right) },  \label{kappa}
\end{equation}%
where $\chi =\partial _{t}$ is the Killing vector. By using the metric (\ref%
{Metric}) and Eq. (\ref{kappa}), we get the surface gravity as%
\begin{equation}
\kappa =\frac{\left. \left( \frac{df\left( r\right) }{dr}\right) \right\vert
_{r=r_{+}}H(\varepsilon )}{2F(\varepsilon )},  \label{k}
\end{equation}%
and by considering the obtained metric function (\ref{f(r)Uch}), Eq. (\ref{m}%
) and the surface gravity (\ref{k}) within the Hawking temperature relation (%
$T_{H}=\frac{\kappa }{2\pi }$), we get it%
\begin{equation}
T_{H}=\frac{H(\varepsilon )\mathcal{B}_{3}A^{2}r_{+}^{4}+3H(\varepsilon
)\left( 1-\mathcal{B}_{2}r_{+}^{2}\right) }{12\pi r_{+}\left(
A^{2}r_{+}^{2}-1\right) F\left( \varepsilon \right) },  \label{Tfinal}
\end{equation}%
where $\mathcal{B}_{n}=$ $\frac{\Lambda }{H^{2}(\varepsilon )}+nA^{2}$.
Also, $r_{+}$ is related to the event horizon of the black hole. The
obtained Hawking temperature depends on all the parameters of these black
holes.

\begin{figure}[tbph]
\centering
\includegraphics[width=0.3\linewidth]{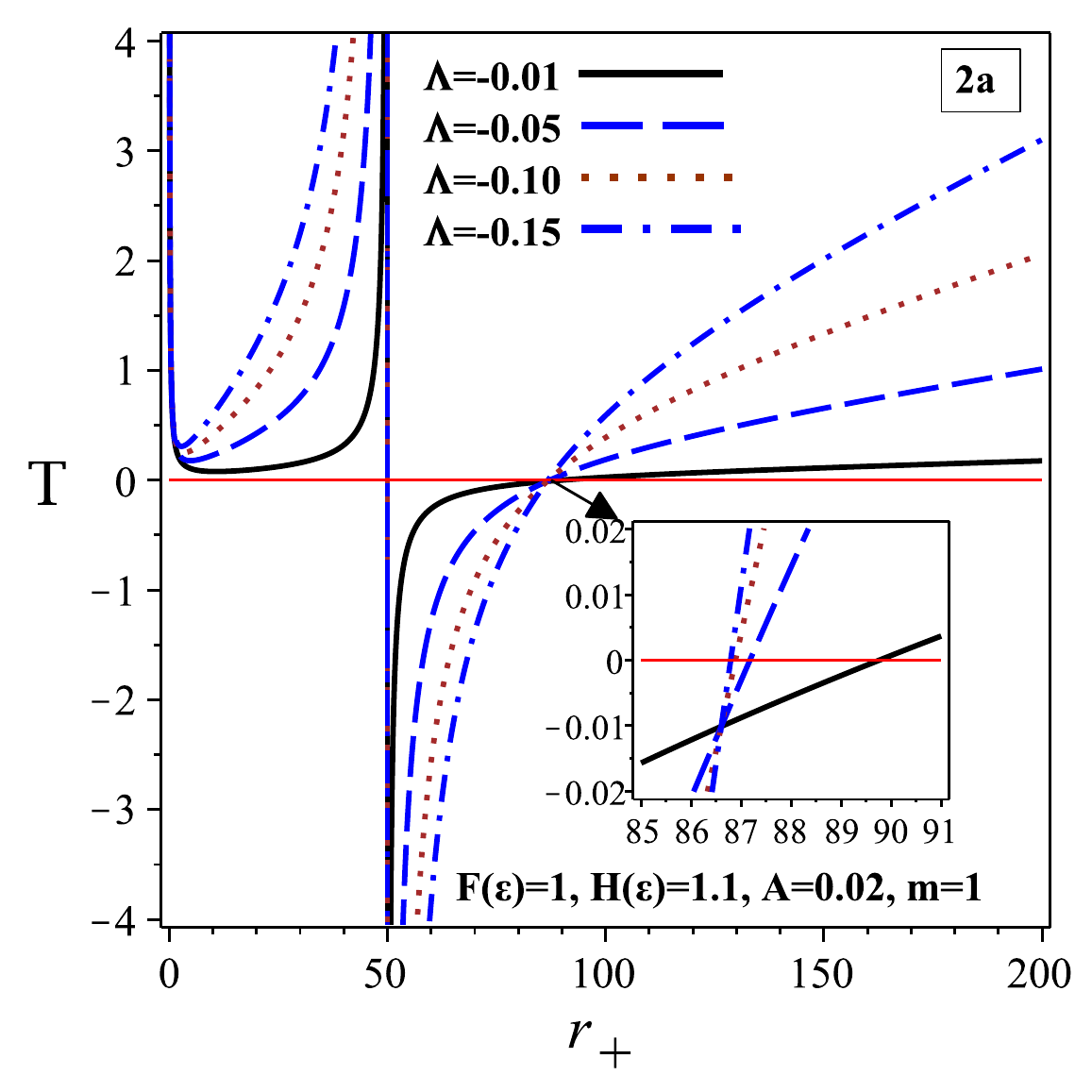} \includegraphics[width=0.3%
\linewidth]{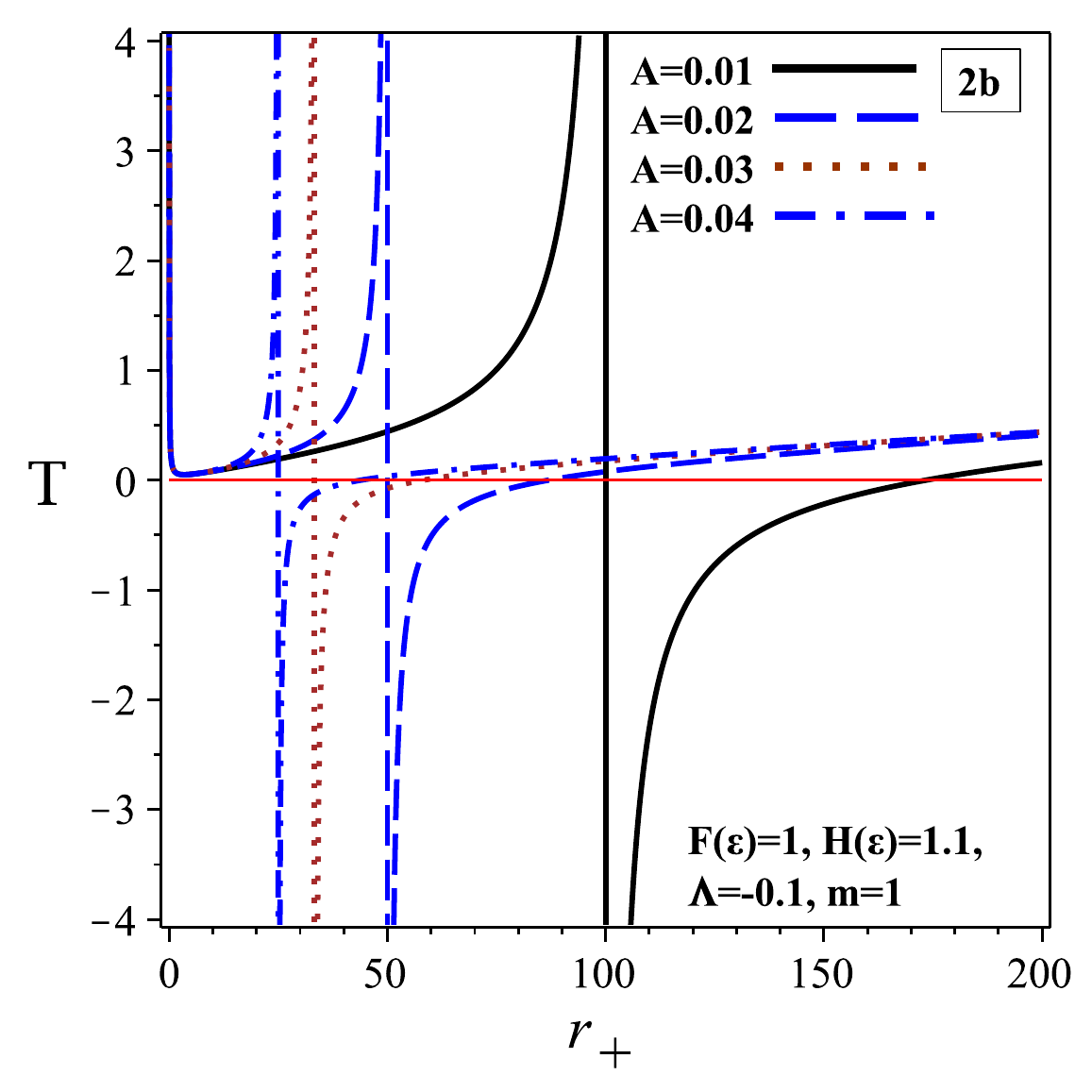} \newline
\includegraphics[width=0.3\linewidth]{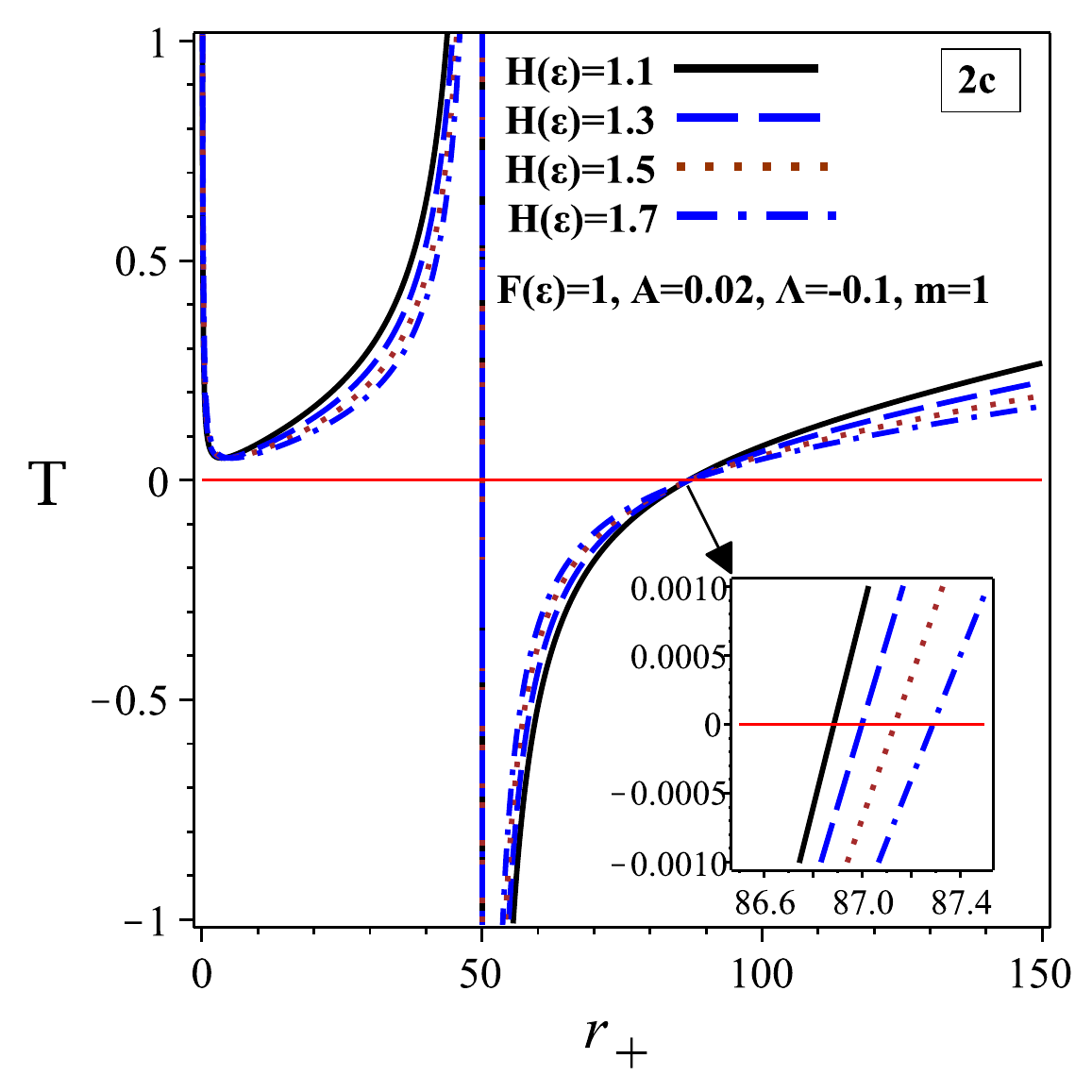} \includegraphics[width=0.3%
\linewidth]{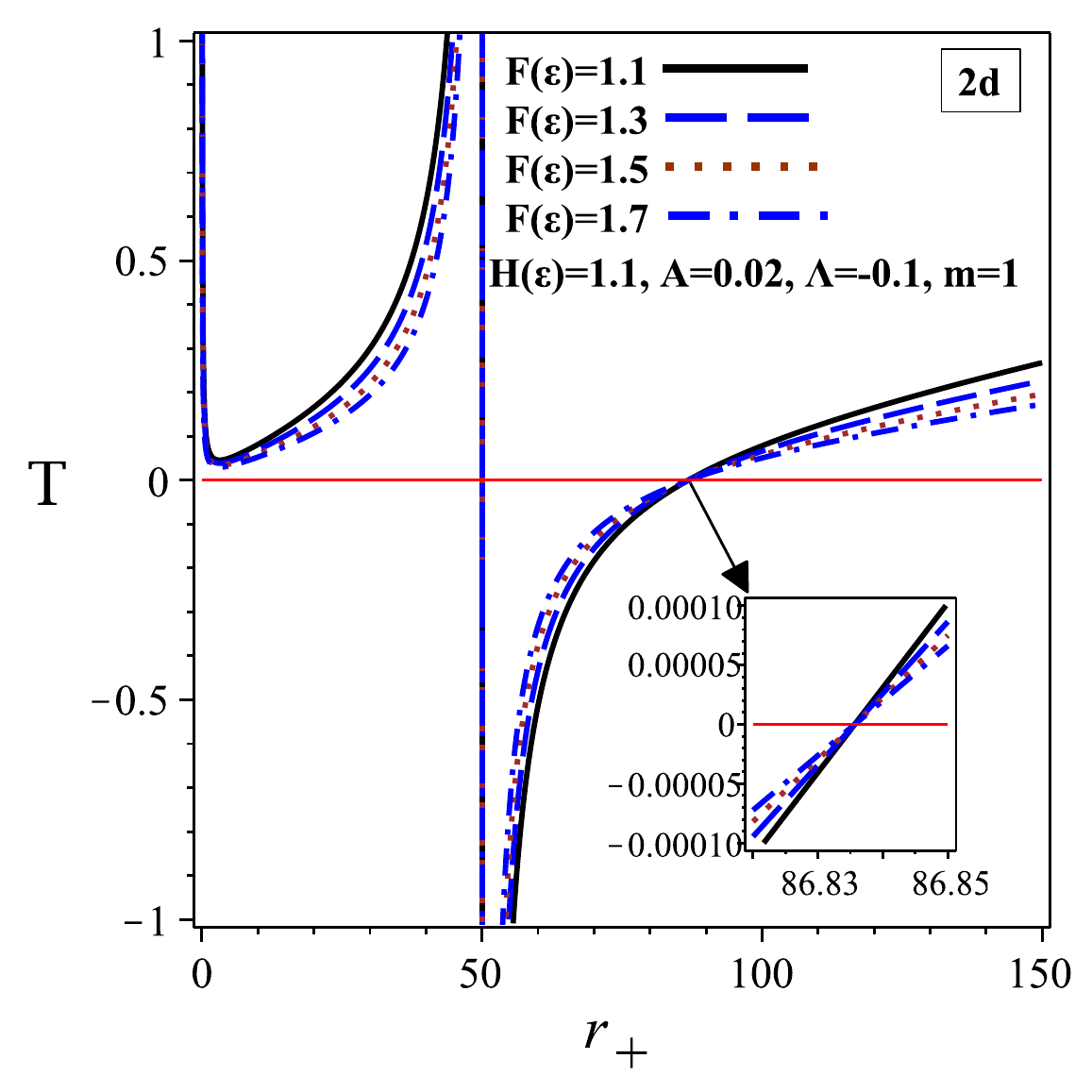} \newline
\caption{$T_{H}$ versus $r_{+}$ for different values of parameters. }
\label{Fig2}
\end{figure}

The high-energy limit (where the limit $r_{+}\rightarrow 0$ is known as the
high-energy limit) of the temperature (\ref{Tfinal}) is given by $\underset{%
r_{+}\rightarrow 0}{\lim }T_{H}\propto \frac{H(\varepsilon )}{4\pi
F\left(\varepsilon \right) r_{+}}$, where indicates that the high-energy
limit of the temperature only depends on rainbow functions $F\left(
\varepsilon\right) $ and $H(\varepsilon )$. Also, in this limit, the
temperature is always positive. As a result, the temperature of small
accelerating black holes in gravity's rainbow is positive.

The asymptotic behavior of the temperature (\ref{Tfinal}) is obtained $%
\underset{r_{+}\rightarrow \infty }{\lim }T_{H}\propto \frac{-H(\varepsilon )%
\mathcal{B}_{3}r_{+}}{12\pi F\left( \varepsilon \right) }$, where reveals
the asymptotic behavior of the temperature is dependent on the cosmological
constant, acceleration parameter and rainbow functions $F\left( \varepsilon
\right) $ and $H(\varepsilon )$. To have the positive temperature, we must
respect to $\mathcal{B}_{3}=\frac{\Lambda }{H^{2}(\varepsilon )}+3A^{2}<0$,
which leads to $\Lambda <-3A^{2}H^{2}(\varepsilon )$. In other words, the
asymptotic behavior of the temperature can be positive when the cosmological
constant is negative.

We plot the Hawking temperature versus $r_{+}$ to study of the behavior of temperature. The results reveal that:

i) there is a singularity for the temperature at $r_{div}=\frac{1}{A}$. This singularity only depends on the accelerating parameter. Such behavior exists for entropy (see Eq. (\ref{S}), in the next subsection). We have to omit this point to have the physical behavior of the temperature and the entropy.

ii) there is a zero point for the temperature ($r_{+_{T=0}}$) which depends on the parameters of the system.

iii) the temperature is positive for $r_{+}<r_{div}$ (small black
holes), and $r_{+}>r_{+_{T=0}}$ (large black holes). 

iv) Fig. \ref{Fig2}a reveals that the roof of $T_{H}$ decreases by
increasing $\left\vert A\right\vert $.

v) Fig. \ref{Fig2}b indicates that by increasing $A$, the divergence point decreases (as we expected), and the root of the temperature decreases.

vi) by increasing $H(\varepsilon )$, the root of temperature
increases (see Fig. \ref{Fig2}c).

vii) the root of temperature is independent of $F\left( \varepsilon
\right) $, see Fig. \ref{Fig2}d. 

\subsection{\textbf{Entropy}}

To obtain the entropy of black holes, one can use the area law in the form $%
S=\frac{\mathcal{A}}{4}$, where $\mathcal{A}$\ is the horizon area and is
defined as 
\begin{equation}
\mathcal{A}=\left. \int_{0}^{2\pi }\int_{0}^{\pi }\sqrt{g_{\theta \theta
}g_{\varphi \varphi }}\right\vert _{r=r_{+}}=\left. \frac{4\pi r^{2}}{%
H^{2}\left( \varepsilon \right) \left( 1-A^{2}r^{2}\right) K}\right\vert
_{r=r_{+}}=\frac{4\pi r_{+}^{2}}{H^{2}\left( \varepsilon \right) \left(
1-A^{2}r_{+}^{2}\right) K},  \label{A}
\end{equation}%
by replacing the horizon area (\ref{A}) within $S=\frac{\mathcal{A}}{4}$,
the entropy of accelerating black holes in gravity's rainbow is given%
\begin{equation}
S=\frac{\pi r_{+}^{2}}{H^{2}\left( \varepsilon \right) \left(
1-A^{2}r_{+}^{2}\right) K},  \label{S}
\end{equation}%
where in the absence of acceleration parameter and rainbow function it turns
to $S=\pi r_{+}^{2}$, as we expected. Indeed, in the absence of the
accelerating parameter, $A$ is zero, and $K=1$.

In addition, there is a singularity for the entropy (\ref{S}) at $r_{+}=%
\frac{1}{A}$. This singularity depends on the acceleration parameter. There
is such behavior for the mass (\ref{m}) and the obtained temperature (\ref%
{Tfinal}). Also, the existence of such singularity is reported for charged
accelerating BTZ black holes \cite{BTZch}. To remove this singularity, we
have to consider $r_{+}\neq \frac{1}{A}$. In other words, we do not permit
to consider $r_{+}=\frac{1}{A}$, because this radius leads to a singularity
in the geometrical mass (\ref{m}), temperature (\ref{Tfinal}) and entropy (%
\ref{S}).

Now, we series the entropy (\ref{S}) versus the event horizon ($r_{+}$) in
the limit $r_{+}\rightarrow 0$ and $r_{+}\rightarrow \infty $, to study the
high-energy and asymptotic behavior of the entropy.

The high-energy limit of the entropy (\ref{S}) is obtained%
\begin{equation}
\underset{r_{+}\rightarrow 0}{\lim }S\propto \frac{\pi r_{+}^{2}}{%
H^{2}\left( \varepsilon \right) K},
\end{equation}%
the high-energy limit of the entropy depends on $H(\varepsilon )$, and $K$.
It is notable that in the high-energy limit, the entropy is positive because 
$K$ and $H^{2}\left( \varepsilon \right) $ are positive. So, the entropy of
small accelerating black holes in gravity's rainbow is always positive.

The asymptotic behavior of the entropy (\ref{S}) is given by%
\begin{equation}
\underset{r_{+}\rightarrow \infty }{\lim }S\propto \frac{-\Lambda }{%
A^{2}H^{2}\left( \varepsilon \right) K},
\end{equation}
where reveals the asymptotic behavior of the entropy can be positive,
provided $\Lambda <0$.

\subsection{\textbf{Heat capacity}}

In the canonical ensemble context, a thermodynamic system's local stability
can be studied by heat capacity. So we study the heat capacity to find the
local stability for such black holes. In other words, we evaluate the
effects of rainbow functions ($F\left( \varepsilon \right) $, and $H\left(
\varepsilon \right) $), and the acceleration parameter $A$ on the local
stability of accelerating black holes.

The heat capacity is defined as 
\begin{equation}
C=\frac{T}{\left( \frac{\partial T}{\partial S}\right) }=\frac{T}{\left( 
\frac{\partial T}{\partial r_{+}}\right) /\left( \frac{\partial S}{\partial
r_{+}}\right) },  \label{Heat}
\end{equation}%
by considering the obtained temperature (\ref{Tfinal}) and the entropy (\ref%
{S}), and some calculations, we can get the heat capacity in the following
form 
\begin{equation}
C=\frac{2\pi \left( A^{2}\mathcal{B}_{3}r_{+}^{4}+3\left( 1-\mathcal{B}%
_{2}r_{+}^{2}\right) \right) }{H^{2}(\varepsilon )\left( A^{6}\mathcal{B}%
_{3}r_{+}^{6}-A^{4}\mathcal{B}_{6}r_{+}^{4}+3\left( \frac{A^{2}\Lambda
r_{+}^{2}}{H^{2}(\varepsilon )}-\mathcal{B}_{-2}-\frac{1}{r_{+}^{2}}\right)
\right) K}.  \label{H}
\end{equation}

To obtain the high-energy limit of the heat capacity (\ref{H}), we evaluate
it in the limit $r_{+}\rightarrow 0$ and get%
\begin{equation}
\underset{r_{+}\rightarrow 0}{\lim }C\propto \frac{-2\pi r_{+}^{2}}{%
H^{2}\left( \varepsilon \right) K},
\end{equation}
where indicates that in the high-energy limit, the heat capacity is
negative. As a result, the heat capacity of small accelerating black holes
in gravity's rainbow is always negative. In other words, the accelerating
black holes with small radius cannot satisfy the local stability condition.

We obtain the asymptotic behavior of the heat capacity (\ref{H}) in the
limit $r_{+}\rightarrow \infty $, which is given by%
\begin{equation}
\underset{r_{+}\rightarrow \infty }{\lim }C\propto \frac{2\pi }{H^{2}\left(
\varepsilon \right) KA^{4}r_{+}^{2}},
\end{equation}
where reveals the asymptotic behavior of the heat capacity is always
positive. Indeed, the accelerating black holes with large radii satisfy the
local stability.

\begin{figure}[tbph]
\centering
\includegraphics[width=0.3\linewidth]{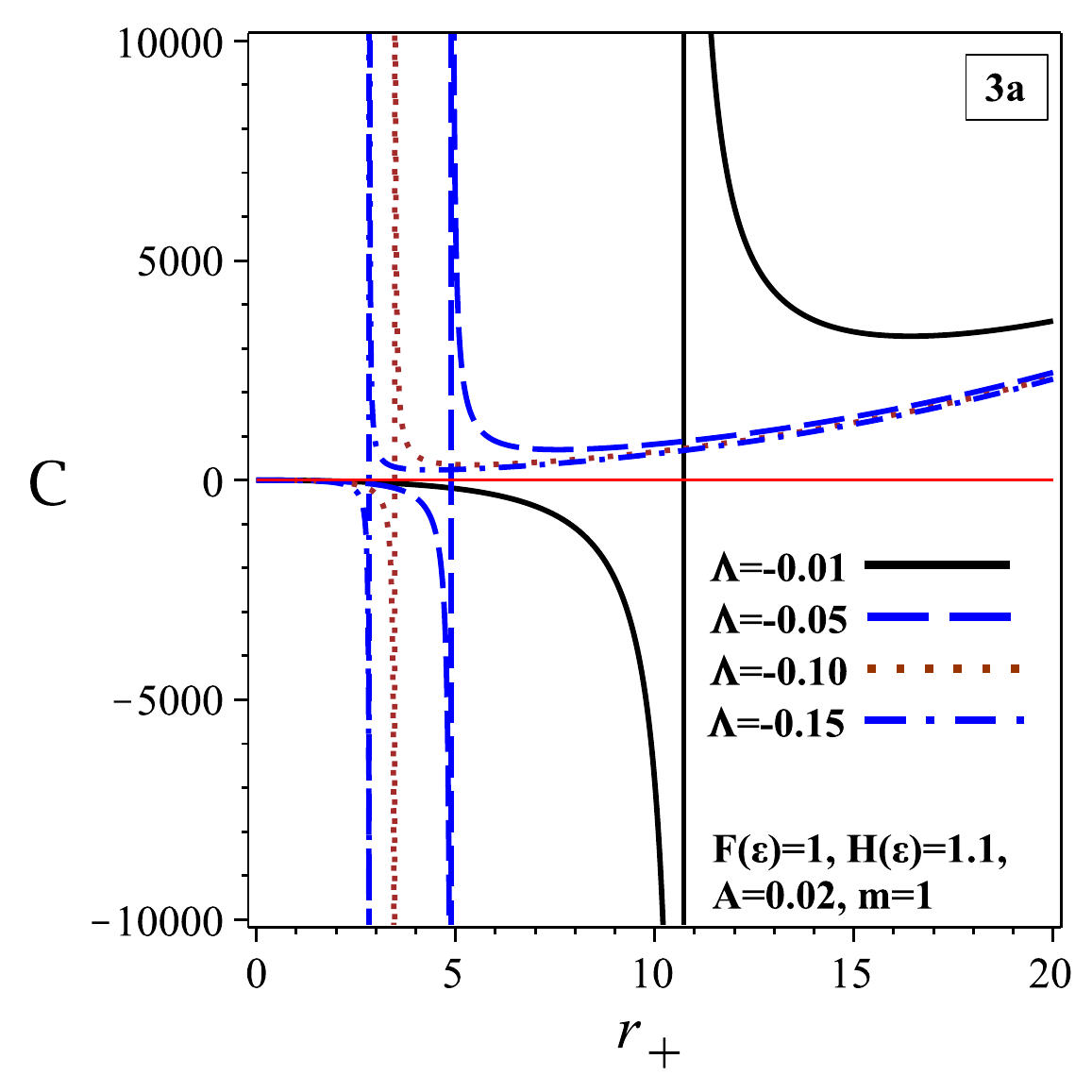} \includegraphics[width=0.3%
\linewidth]{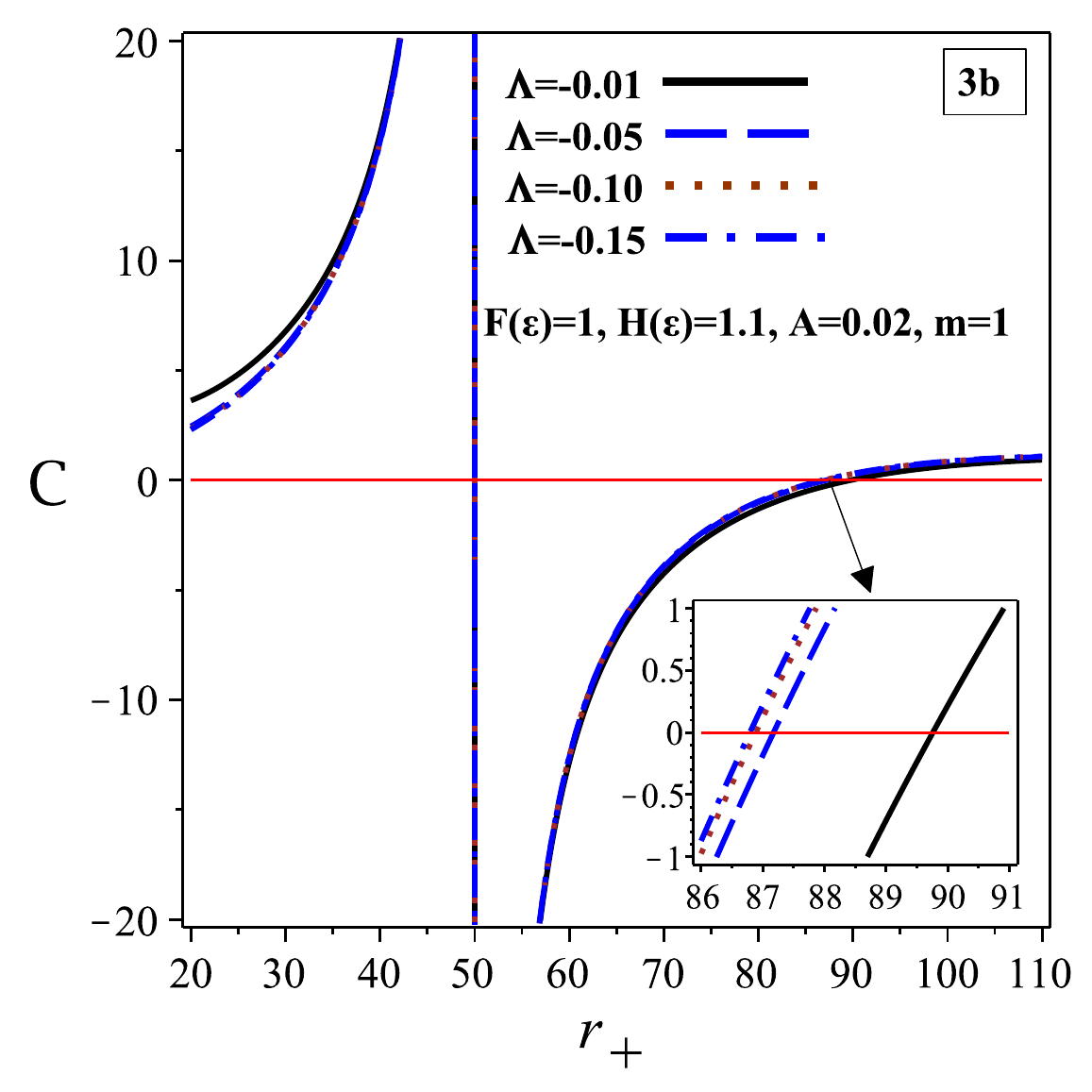} \newline
\includegraphics[width=0.3\linewidth]{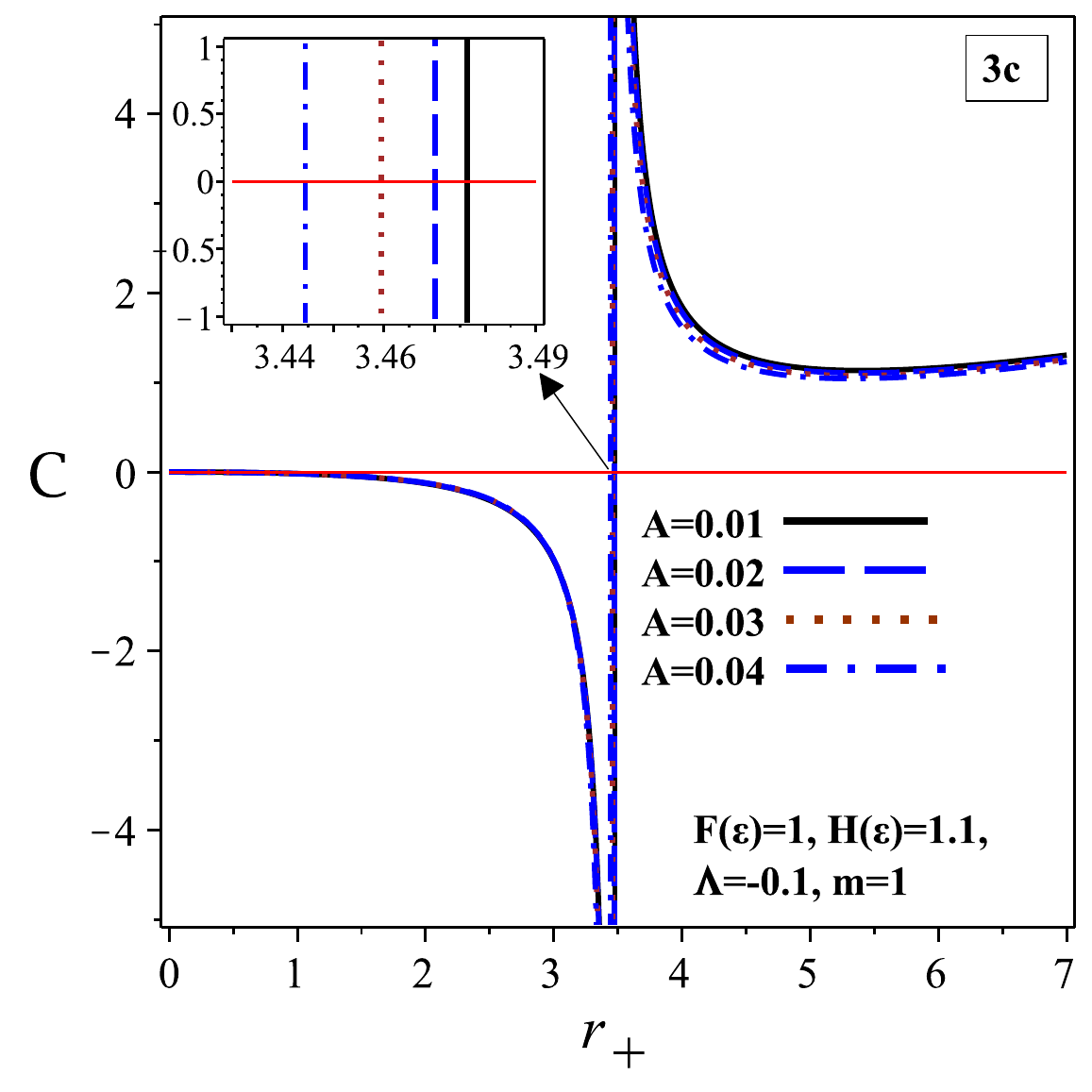} \includegraphics[width=0.3%
\linewidth]{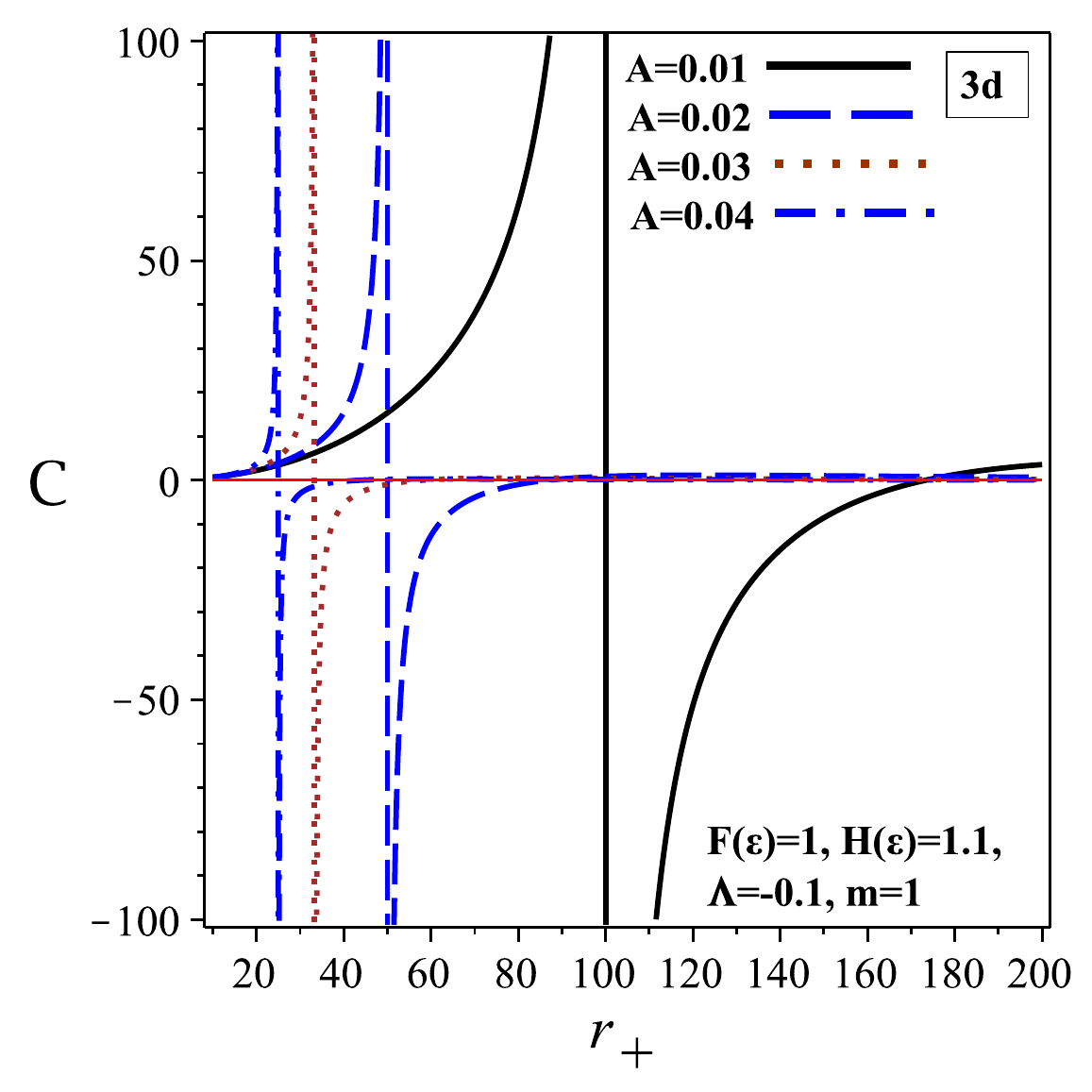} \newline
\includegraphics[width=0.3\linewidth]{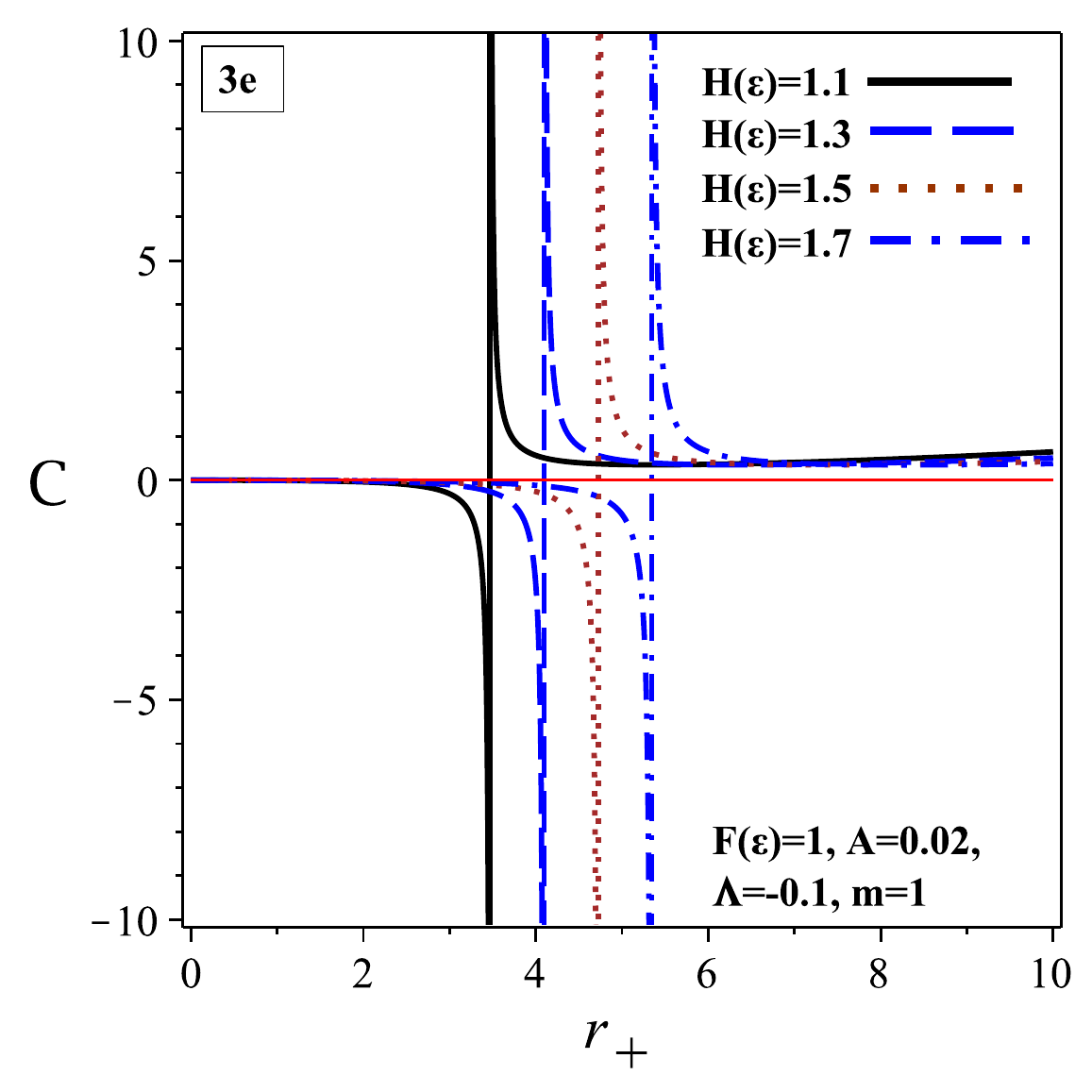} \includegraphics[width=0.3%
\linewidth]{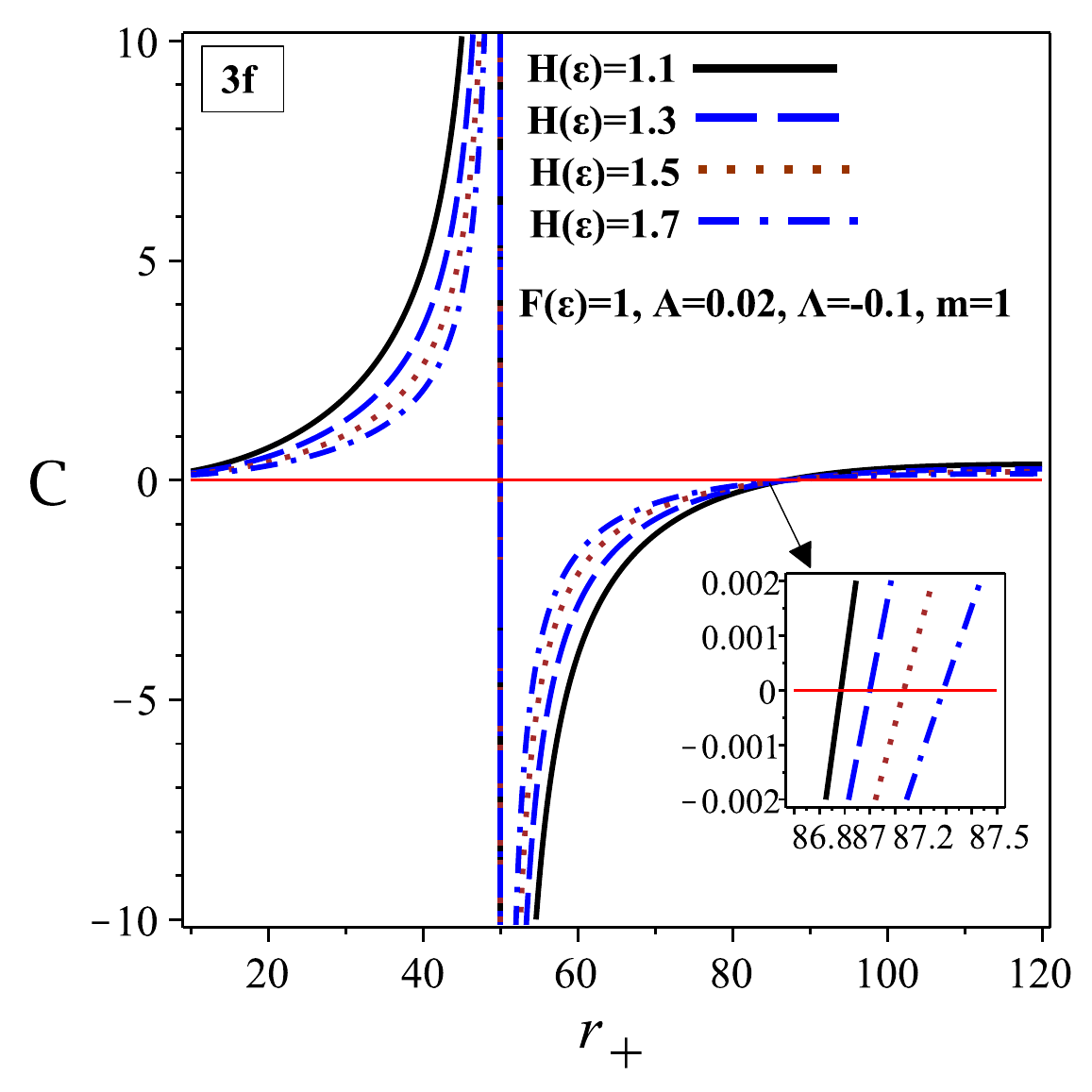} \newline
\includegraphics[width=0.3\linewidth]{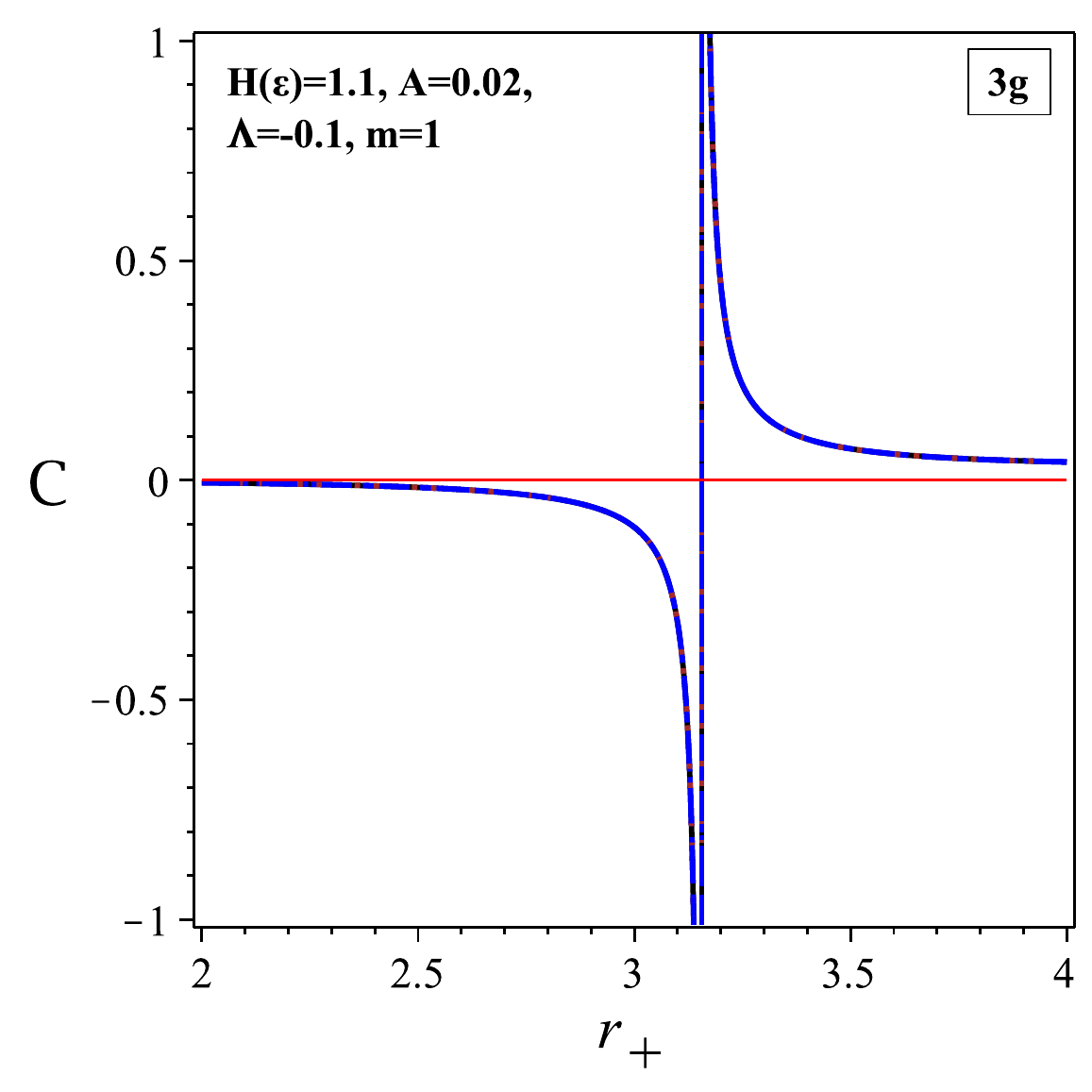} \includegraphics[width=0.3%
\linewidth]{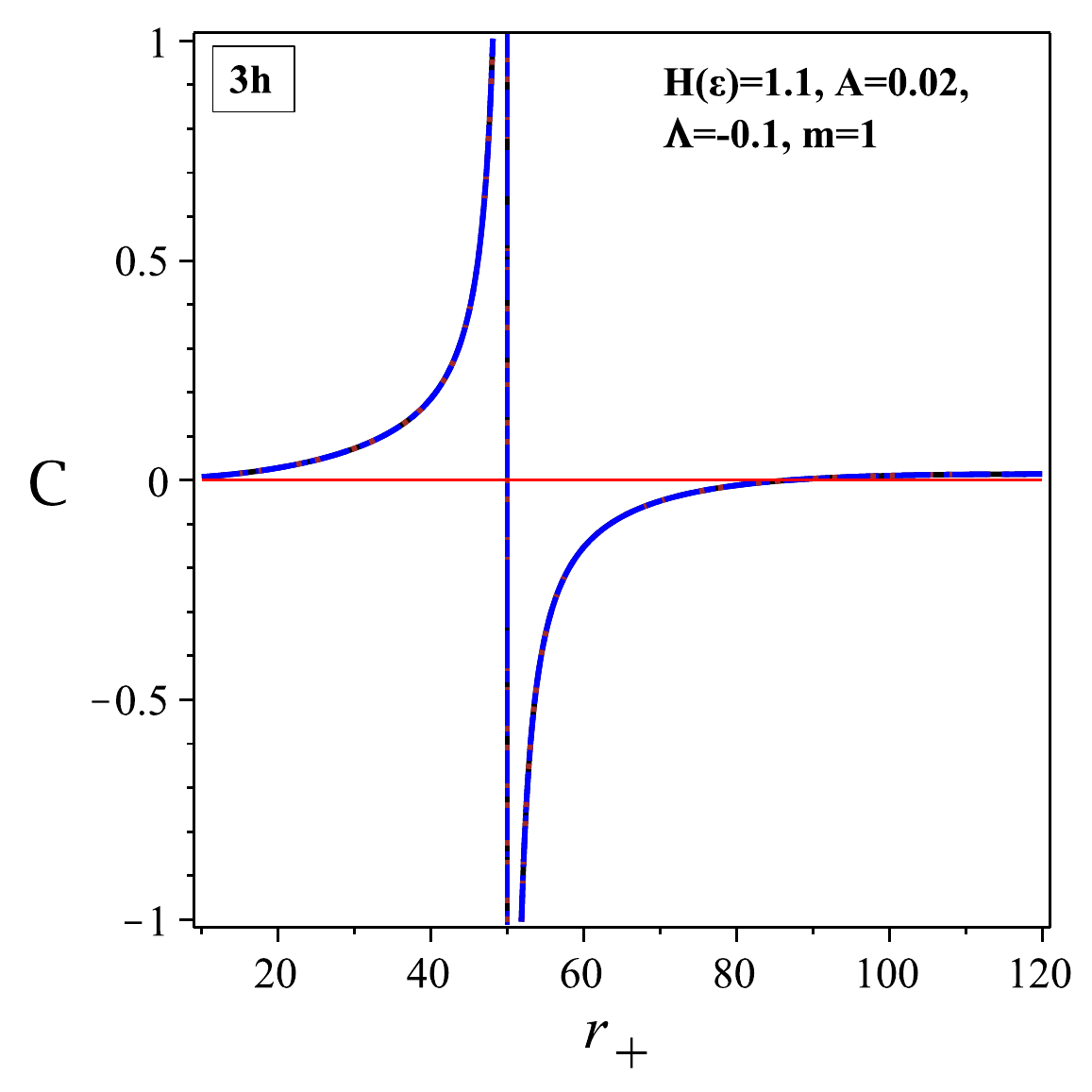} \newline
\caption{$C$ versus $r_{+}$ for different values of parameters. }
\label{Fig3}
\end{figure}

Our analysis reveals two important points, which are; i) the temperature and
entropy of small accelerating black holes are always positive, but the heat
capacity is negative. ii) The large accelerating black holes can have
positive values for the temperature and entropy when the cosmological
constant is negative. In other words, large AdS accelerating black holes are
physical objects. However, the heat capacity is always positive in this
range.

According to the behavior of the temperature, entropy, and heat capacity,
accelerating black holes with large radii in gravity's rainbow can be
physical objects and satisfy the local stability when the cosmological
constant is negative.

We plot the heat capacity versus $r_{+}$ for further investigation. The results show that:

i) there are two divergence points for the heat capacity, which we denote by $r_{{div}_{1}}$, and $r_{{div}_{2}}$. $r_{{div}_{1}}$ refers to the first divergence point and $r_{{div}_{2}}$ refers to the second divergence point. There is also a zero point ($r_{C=0}$). Further details can be found in Fig. \ref{Fig3}.

ii) the heat capacity at $r_{{div}_{1}}<r_{+}<r_{{div}_{2}}$, and $r_{+}>r_{C=0}$, is positive.

iii) by increasing $\left\vert \Lambda \right\vert $, $r_{{div}_{1}}$ decreases but $r_{{div}_{2}}$ does not change (see Figs. \ref{Fig3}a and \ref{Fig3}b). By comparing the temperature (Fig. \ref{Fig2}a) and the heat capacity (Figs. \ref{Fig3}a and \ref{Fig3}b), we find that the local stability area increases by increasing $\left\vert \Lambda \right\vert $.

iv) Figs. \ref{Fig3}c and \ref{Fig3}d, indicate that $r_{{div}_{1}}$, $r_{{div}_{2}}$ and $r_{C=0}$ decrease by increasing the accelerating parameter. In addition, the local stability area increases by increasing $A$ (compare Fig. \ref{Fig2} a with Figs. \ref{Fig3}c and \ref{Fig3}d).

v) by increasing $H(\varepsilon )$, the first divergence point ($r_{{div}_{1}}$), and the root of the heat capacity ($r_{C=0}$) increase. However, the second divergence point ($r_{{div}_{2}}$) of the heat capacity does not change (see Figs. \ref{Fig3}e and \ref{Fig3}f). By comparing the temperature (Fig. \ref{Fig2}c) and the heat capacity (Figs. \ref{Fig3}e and \ref{Fig3}f), the local stability area decreases by increasing $H(\varepsilon )$.

vi) the heat capacity is independent of the change of $F(\varepsilon )$, see Figs. \ref{Fig3}g and \ref{Fig3}h. So, the local stability area does not change by varying $F(\varepsilon )$.

\section{\textbf{Geodesic Equations}}

First, we need to determine the geodesic motion of photons in a rotationally
symmetric and static spacetime metric in the presence of gravity's rainbow.
The modified $C-$metric, which depends on four parameters: the mass
parameter $m$ with length dimension, the acceleration parameter $A$ with
inverse length dimension \cite{Acc4,Acc5}, the cosmological constant $%
\Lambda $ and rainbow functions, is used to describe this scenario. The
rainbow functions, which are introduced in the model, further influence the
motion of photons in this spacetime. To analyze this motion, we formulate
the Lagrangian $\mathcal{L}(x,\dot{x})=\frac{1}{2}\mathrm{g}_{\mu \nu }\dot{x%
}^{\mu }\dot{x}^{\nu }$ for the geodesics. The Lagrangian allows us to
analyze how photons move around an accelerating black hole under the effects
of Gravity's Rainbow can be written as 
\begin{equation}
\mathcal{L}=\frac{1}{2\,\mathcal{K}^{2}\left( r,\theta \right) }\left( -%
\frac{f(r)}{F^{2}\left( \varepsilon \right) }\dot{t}^{2}+\frac{1}{%
f(r)H^{2}\left( \varepsilon \right) }\dot{r}^{2}+\frac{r^{2}}{g\left( \theta
\right) H^{2}\left( \varepsilon \right) }\dot{\theta}^{2}+\frac{g\left(
\theta \right) r^{2}\sin ^{2}\theta }{H^{2}\left( \varepsilon \right) {K^{2}}%
}\dot{\varphi}^{2}\right) ,  \label{LagrangianCM}
\end{equation}%
with the dot notation that represents differentiation with respect to an
affine parameter denoted as $\lambda $, which is defined along the geodesic.
To examine the trajectories of light rays in the vicinity of an accelerating
black hole, as well as determine the radius of the photon sphere and the
corresponding critical impact parameter, it is permissible to confine the
motion of the photon to the equatorial plane of the accelerating black hole,
that is, $\theta =\pi /2$ and $\dot{\theta}=0$. Since the metric coefficient
function cannot be determined by the $t$ and $\theta $ coordinates, there
exist two conserved quantities, $E$, and $L$ \cite%
{CapozzielloJCAP2023,Capozziello2023}, which correspond to energy and
angular momentum, respectively. The given expressions are 
\begin{equation}
E=-\frac{\partial \mathcal{L}}{\partial \dot{t}}=\frac{f(r)}{F^{2}\left(
\varepsilon \right) }\dot{t},~~\&~~L=\frac{\partial \mathcal{L}}{\partial 
\dot{\varphi}}=\frac{r^{2}\dot{\varphi}}{H^{2}\left( \varepsilon \right) }.
\label{ConstMotions}
\end{equation}

The orbit equation for the null geodesic $\mathcal{L}=0$ can thus be found
as follows 
\begin{equation}
\left( \frac{dr}{d\varphi }\right) ^{2}=r^{2}f(r)\left( \frac{%
r^{2}F^{2}(\varepsilon )}{b^{2}f(r)H^{2}(\varepsilon )}-1\right) .
\label{OrbitEq1}
\end{equation}

We can observe that the term on the right-hand side of the equation serves
as an effective potential for particles moving in the $r$ direction. Thus,
it is evident that the orbit equation for a particular metric relies solely
on one constant of motion, such as the impact parameter $b=L/E$. 
In cases where the light ray approaches the center and then exits after
reaching a minimum radius $R$, it is more convenient to express the orbit
Eq. (\ref{OrbitEq1}) in terms of $R$ rather than $b$. Since $R$ represents
the turning point of the trajectory, the condition $dr/d\varphi |_{R=0}$
must be satisfied \cite{PerlickPR2022}. By using the orbit Eq. (\ref%
{OrbitEq1}), we can derive the relationship between $R$ and the impact
parameter $b$ as 
\begin{equation}
\frac{1}{b^{2}}=\frac{f(R)H^{2}(\varepsilon )}{R^{2}F^{2}(\varepsilon )}.
\label{R&b}
\end{equation}

We shall now introduce the following function 
\begin{equation}
U(r)=\sqrt{\frac{r^{2}F^{2}(\varepsilon )}{f(r)H^{2}(\varepsilon )K^{2}}}.
\label{Ur}
\end{equation}

It is useful to define the impact parameter $b$ with respect to the function 
$U(r)$ as $b\equiv U(R)$. When the acceleration parameter and gravity's
rainbow are absent as well as $\Lambda =0$, the function $U(r)$ can be
considered equivalent to the \textquotedblleft effective potential" found in
the Schwarzschild case. This effective potential pertains to the motion of
photons in Schwarzschild's gravity. By substituting Eqs. (\ref{R&b}) and (%
\ref{Ur}) in Eq. (\ref{OrbitEq1}), we can re-express Eq. (\ref{OrbitEq1}) in
the following form 
\begin{equation}
\left( \frac{dr}{d\varphi }\right) ^{2}=\frac{r^{2}f(r)}{K^{2}}\left( \frac{%
U^{2}(r)}{U^{2}(R)}-1\right) .  \label{OrbitEq2}
\end{equation}

Likewise, we can safely consider the limit $R \rightarrow r_{\text{ph}}$
from this point forward.

To proceed, it is necessary to determine the radius of the photon sphere,
denoted as $r_{\text{ph}}$. This can be accomplished by satisfying two
conditions along a circular light orbit; $V_{\text{eff}}|_{r=r_{\text{ph}%
}}=0 $ and $\frac{dV_{\text{eff}}}{dr}|_{r=r_{\text{ph}}}=0$. By solving
these two equations simultaneously, we obtain the equation for the radius of
a circular light orbit, which takes the form 
\begin{equation}
\frac{f^{\prime }(r_{\text{ph}})}{f(r_{\text{ph}})}-\frac{2}{r_{\text{ph}}}%
=0.  \label{Cond12Simul-2}
\end{equation}%
Eqs. (\ref{Ur}) and (\ref{Cond12Simul-2}) allow us to obtain both the
position of the photon sphere and the critical impact parameter, $b_{\text{c}%
}\equiv U\left( r_{\text{ph}}\right) $, given by 
\begin{equation}
r_{\text{ph}}=\frac{\sqrt{\eta _{3}}-1}{2A^{2}m},~~\,\&\,\,~b_{\text{c}}=%
\frac{\sqrt{3\,\eta _{2}}F(\varepsilon )}{\sqrt{3A^{2}\eta _{1}-\eta
_{2}\Lambda }},
\end{equation}%
where 
\begin{eqnarray}
\eta _{1} &=&\sqrt{\eta _{3}}-1-8A^{4}m^{4}\left( 1+\frac{H^{2}(\varepsilon
)\left( 5-2\sqrt{\eta _{3}}\right) }{4A^{2}m^{2}}\right) ,  \notag \\
&&  \notag \\
\eta _{2} &=&1-\sqrt{\eta _{3}}+3A^{2}m^{2}\left( 3-\sqrt{\eta _{3}}\right) ,
\notag \\
&&  \notag \\
\eta _{3} &=&1+12A^{2}m^{2}.
\end{eqnarray}

We now aim to explore the trajectory of a light ray as it traverses the
vicinity of a black hole. The transformation $x=\frac{1}{r}$ is a convenient
choice \cite{GrallaPRD2019,PengCPC2021}. This results in the transformation
of the orbit equation as follows 
\begin{equation}
\left( \frac{dx}{d\phi }\right) ^{2}=W(x),  \label{OEq2}
\end{equation}%
with 
\begin{equation}
W\left( x\right) =\left( \frac{F^{2}\left( \varepsilon \right) }{%
b^{2}H^{2}\left( \varepsilon \right) }-\left( x^{2}-A^{2}\right) \left(
1-2mx\right) +\frac{\Lambda }{3H^{2}\left( \varepsilon \right) }\right) .
\end{equation}

Regarding the interaction of light rays with a black hole: i) In the case
when the impact parameter $b>b_{c}$, the light beam approaches one nearest
point before retreating back to infinity from the black hole. ii) The light
ray infalls into the black hole in all cases when the impact parameter is $%
b<b_{c}$. iii) At a radius of $r_{c}$, or the photon sphere radius, the
light ray will circle the black hole when $b=b_{c}$ \cite%
{GrallaPRD2019,PengCPC2021}.

The smallest positive real root of $W(x)=0$, which we indicate as $x_{m} $,
is the turning point for the $b>b_{c}$ case. Using Eq. (\ref{OEq2}), one may
compute 
\begin{equation}
\varphi =2\int_{0}^{x_{m}}\frac{dx}{\sqrt{W(x)}},\quad b>b_{c},
\end{equation}%
to find the whole variation in azimuthal angle $\varphi $ for a given
trajectory with impact parameter $b$. In the case where $b<b_{c}$, our
interest is directed towards the trajectory beyond the horizon $r_{h}$.
Consequently, the overall variation in the azimuthal angle $\varphi $ may be
determined using the integration%
\begin{equation}
\varphi =\int_{0}^{x_{+}}\frac{dx}{\sqrt{W(x)}},\quad b<b_{c},
\end{equation}%
where $x_{+}=\frac{1}{r_{+}}$.

The authors in [18] classified trajectories into three categories, direct,
lensed, and photon rings to analyze the apparent characteristics of emission
emanating from close proximity to a black hole. We provide a brief overview
here. The total number of orbits $n$, given by $n=\frac{\varphi }{2\pi }$,
is dependent on the impact parameter $b$. We express the solution of $n(b)=%
\frac{2\xi -1}{4}$, where $\xi =1,2,3,\dots $ by $b_{\xi}^{\pm }$ so that $%
b_{\xi }^{-}<b_{c}$ and $b_{\xi }^{+}>b_{c}$. Following this, all
trajectories can be categorized in this way: i) direct rays case corresponds
to $\frac{1}{4}<n<\frac{3}{4}$, with $b\in \left(b_{1}^{-},b_{2}^{-}\right)
\cup \left(b_{2}^{+},\infty \right) $. ii) lensing rings case corresponds to 
$\frac{3}{4}<n<\frac{5}{4}$, with $b\in \left( b_{2}^{-},b_{3}^{-}\right)
\cup \left(b_{3}^{+},b_{2}^{+}\right) $. iii) photon rings case corresponds
to $n>\frac{5}{4}$, with $b\in \left(b_{3}^{-},b_{3}^{+}\right) $ \cite%
{GrallaPRD2019,PengCPC2021}.

\begin{figure}[tbph]
\centering
\includegraphics[width=0.3 \linewidth]{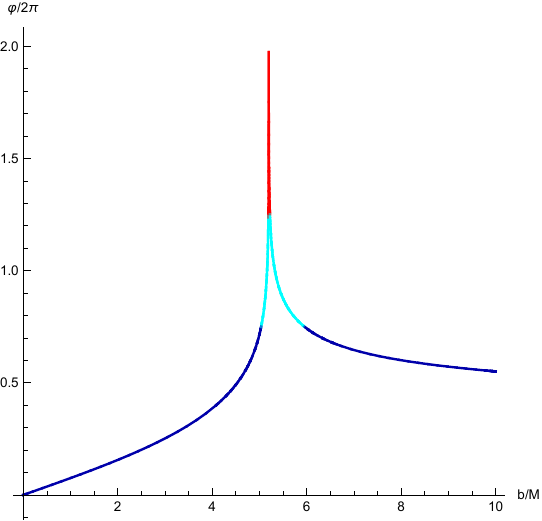} %
\includegraphics[width=0.3 \linewidth]{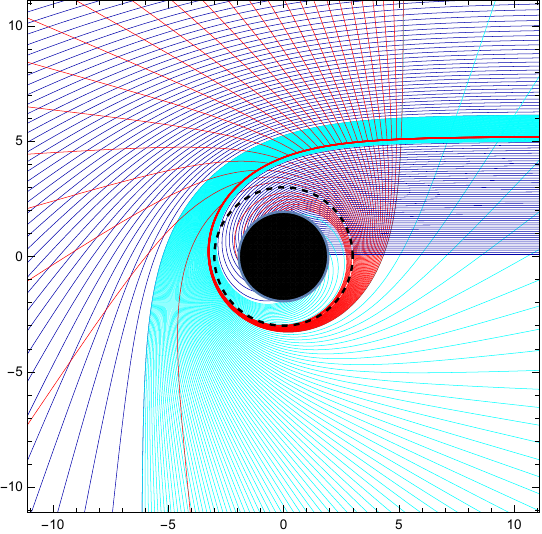} \newline
\includegraphics[width=0.3 \linewidth]{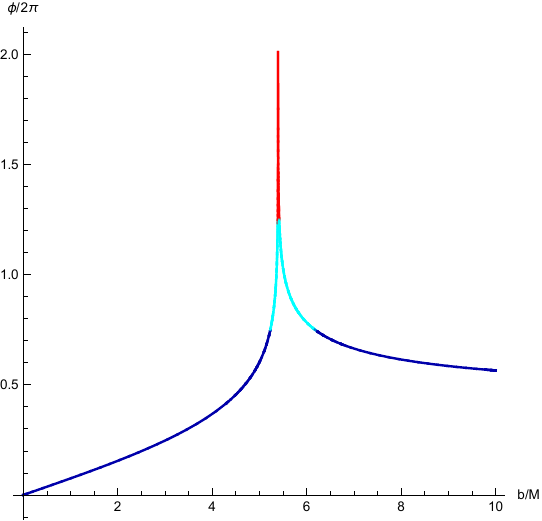} %
\includegraphics[width=0.3 \linewidth]{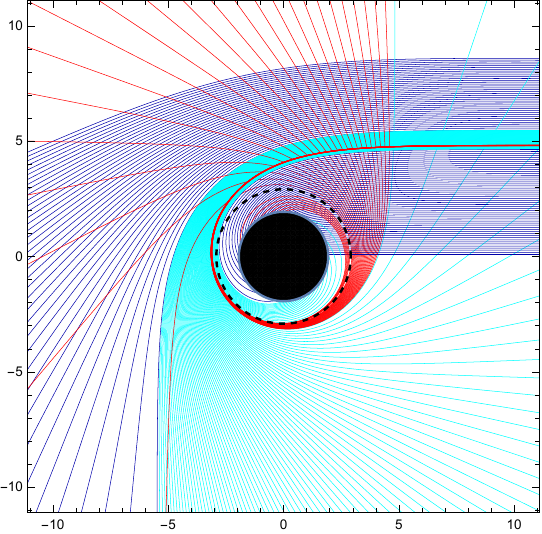} \newline
\caption{Photon behavior for various values of $A=0.001, 0.1$. A selection
of related photon trajectories is displayed in the right panel, taking $(r, 
\protect\varphi)$ as Euclidean polar coordinates, while the fractional
number of orbits, $n = \frac{\protect\varphi}{2\protect\pi}$, is displayed
in the left panel. $\protect\varphi$ represents the total change in (orbit
plane) azimuthal angle outside the horizon.}
\label{Fig4}
\end{figure}

In Fig. \ref{Fig4}, one can observe the variations in the trajectory of
light rays by changing both the acceleration parameter $A$ and impact
parameter $b$. As such, we show how the trajectory of light rays changes for
associated black hole parameters, including the acceleration parameter,
while keeping fixed values for the cosmological constant $\Lambda$ and the
rainbow functions $F(\varepsilon)$ and $H(\varepsilon)$. Figure \ref{Fig2}
depicts the number of orbits $n$ versus the impact parameter $b$. The blue
color indicates direct emission rays, cyan denotes lensing rays, and red
denotes photon ring rays. The photon orbit and the event horizon of the
black hole are depicted by the dashed black circle and the black disk
respectively in the ray tracing picture.

Moreover, if we set $m=1$, $\Lambda =-0.02$, $F(\varepsilon)=1$, $%
H(\varepsilon)=0.9$, we can see from the table and figures that the range of
lensing rings grows with the increasing acceleration parameter $A$. In the $%
(b,\varphi )$ plane, the photon orbit displays a narrow peak when the impact
parameter is extremely close to the critical impact parameter $b\pm b_{c}$ 
\cite{UniyalPoDU2023}. Afterward, as $b$ grows, the photon trajectories are
always direct rays in any scenario.

\begin{table}[ht]
\caption{Regions of direct rays, lensing rings, and photon rings for two
different values of the acceleration parameter $A$.}
\label{table1}\centering
\begin{tabular}{|c|c|c|}
\hline\hline
Acceleration parameter & $A=0.001$ & $A=0.1$ \\[0.5ex] \hline
Direct rays & $b<5.04$ & $b<5.23$ \\ 
$\left(n < \frac{3}{4}\right)$ & $b>5.95$ & $b>6.18$ \\ \hline
Lensing rings & \quad$5.04<b<5.19$ & \quad $5.23<b<5.39$ \\ 
$\left(\frac{3}{4}<n<\frac{5}{4}\right)$ & \quad $5.22<b<5.95$ & \quad $%
5.42<b<6.18$ \\ \hline
Photon ring $\left(n>\frac{5}{4}\right)$ & \quad $5.19<b<5.22$ & \quad $%
5.39<b<5.42$ \\[1ex] \hline
\end{tabular}%
\end{table}

\section{\textbf{Discussion and Conclusions}}

We extracted accelerating black hole solutions in gravity's rainbow. Then,
we studied the effects of various parameters on the horizons of these black
holes in Fig. \ref{Fig1}. We evaluated the Hawking temperature and entropy
for these black holes to find the local stability area using heat capacity.
Our findings indicated that the accelerating black holes with large radii
satisfied the local stability. In addition, we found that there were two divergence points and one real root for the heat capacity. The local stability areas increased by increasing $\left\vert \Lambda \right\vert $, and the accelerating parameter ($A$). However, the local stability decreased by increasing $H(\varepsilon)$. It is worthwhile to mention that the local stability was independent of another rainbow function ($F(\varepsilon)$).

The near-horizon region of energy-dependent $C-$metric is influenced by both
the acceleration parameter and rainbow functions of the AdS black hole,
making it an interesting area to explore. Of particular interest is the
strong gravitational lensing effect in this region. A critical curve with an
impact parameter $b_{c}$ which gives rise to a photon sphere with a radius $%
r_{\text{ph}}$, provides a valuable opportunity to investigate the
trajectories of light rays categorized as direct, lensed, and photon rings.

As a consequence, given the dependence of the event horizon and critical
impact parameter on the rainbow functions (noting that, for the event
horizon, this dependence is confined to $H(\varepsilon)$), one observes that
an increase in the values of $H(\varepsilon)$ results in an increase and
decrease, respectively, in the event horizon and critical impact parameter.
Meanwhile, the photon sphere radius remains unaffected by these functions.

Moreover, our findings indicate a dependency of the critical impact
parameter, photon sphere radius, and event horizon on the acceleration
parameter. In turn, both the photon sphere radius and event horizon exhibit
a similar trend in response to changing acceleration parameter values,
whereas the trend differs for the critical impact parameter. Specifically,
an increase in the acceleration parameter leads to a decrease in both the
photon sphere radius and event horizon, while causing an increase in the
critical impact parameter.

Besides, we observed that the range of lensing rings increased with the
rising acceleration parameter $A$. In the $(b, \varphi)$ plane, the photon
orbit exhibited a narrow peak when the impact parameter was near the
critical value. Subsequently, as $b$ increased, the photon trajectories
consistently presented as direct rays in all scenarios.

\acknowledgements{We are grateful to the anonymous referees for the insightful comments and suggestions, which have allowed us to improve this paper significantly. B. Eslam Panah thanks the University of Mazandaran.
The work of S. Zare and H. Hassanabadi was supported by the Long-Term Conceptual Development of a University of Hradec Kr\'alov\'e for 2023, issued by the Ministry of Education, Youth, and Sports of the Czech Republic.}

\end{document}